\documentclass[11pt]{article}
\usepackage{epsfig, amsmath, amssymb, amsthm, times}
\usepackage{graphicx}
\newtheorem {thm}{Theorem}[section]
\newtheorem {lem}[thm]{Lemma}

\newtheorem {cor}[thm]{Corollary}

\theoremstyle{defintion}
\newtheorem {df}[thm]{Definition}

\theoremstyle{remark}

\theoremstyle{example}
\newtheorem{ex}[thm]{Example}

\theoremstyle{assumption}

\def\pf{{\it Proof.\;}}

\def\R{{\mathbb R}}
\def\N{{\mathbb N}}

\def\Z{{\mathbb Z}}

\DeclareMathOperator{\Cay}{Cay}
\renewcommand\ln{\operatorname{ln}}
\newcommand{\iu}{{i\mkern1mu}}

\providecommand{\keywords}[1]{\textbf{\textit{Key words:}} #1}
\providecommand{\subjclass}[1]{\textbf{\textit{AMS subject classifications:}} #1}

\def\lbl{\label}
\def\be{\begin{equation}}
\def\ee{\end{equation}}

\def\qed{\square}
\def\t{\intercal}
\def\1{\mathbf{1}}

\def\vx{\mathbf{x}}
\def\vy{\mathbf{y}}
\def\vf{\mathbf{f}}
\def\vg{\mathbf{g}}
\def\vv{\mathbf{v}}

\title{Synchronization of coupled chaotic maps
}
 \author{
 Georgi S. Medvedev  and Xuezhi Tang
\thanks{
Department of Mathematics, Drexel University, 3141 Chestnut Street,
Philadelphia, PA 19104, {\tt medvedev@drexel.edu}, {\tt  xt32@drexel.edu} 
}
}

\begin{document}
\maketitle
\begin{abstract}
We prove a sufficient condition for synchronization for coupled one-dimensional
maps and estimate the size of the window of parameters where
synchronization takes place. It is shown that coupled systems on graphs
with positive eigenvalues (EVs) of the normalized graph Laplacian concentrated
around $1$ are more amenable for synchronization. In the light of this condition,
we review spectral properties of Cayley, quasirandom, power-law graphs, and expanders
and relate them to synchronization of the corresponding networks. The analysis of
synchronization on these graphs is illustrated with numerical experiments.
The results of this paper highlight 
the advantages of random connectivity for
synchronization of coupled chaotic dynamical systems.   
\end{abstract}
\\

\section{Introduction}
Among problems concerning dynamics of large networks, synchronization
holds a special place \cite{Strogatz-Sync, PikRos-Sync}. 
With applications ranging from the genesis of epilepsy \cite{JirCur13}
to stability of power grids \cite{MotMye13}, understanding principles underlying synchronization
in real world networks is of the utmost importance. From the theoretical standpoint, the mathematical
analysis of synchronization provides valuable insights into the role of the network topology 
in shaping collective dynamics. 

The mechanism of synchronization in coupled dynamical systems in general depends on the type 
of dynamics generated by individual subsystems. For many diffusively coupled systems, the contribution
of network organization to synchronization can be effectively described by the smallest positive
eigenvalue of the graph Laplacian,
also called the algebraic connectivity of the network \cite{Fied73}.
This has been shown for coupled scalar differential equations (so-called, consensus protocols) 
\cite{OlfMur04, OlfFax07}, limit cycle oscillators \cite{Med11}, excitable systems forced by 
noise \cite{MZ12}, and certain slow-fast systems \cite{MZ12a}. Interestingly, stochastic stability
of the synchronous state can be estimated in terms
of the total effective resistance of the graph \cite{Med12, YouSca10}. Synchronization of the coupled
chaotic systems features a new effect. It turns out that the parameter domain for synchronization
depends on both the smallest and the largest positive eigenvalues of the graph Laplacian.
This has been observed for
systems of coupled differential equation models in \cite{PecCar90, NisMot10}
and for coupled map lattices \cite{JosJoy02}. Furthermore, the analysis in \cite{JosJoy02} shows
that coupled systems on graphs, whose graph Laplacians have all  positive 
EVs localized, are optimal for synchronization
in the sense that they synchronize in a broader range of parameters compared to the systems
on the graphs whose eigenvalues are spread out. The conclusions in \cite{JosJoy02} are based
on formal linear stability analysis of synchronous solutions of coupled systems of chaotic maps.

The goal of the present paper is twofold. First, we rigorously prove a sufficient condition for 
synchronization for systems of coupled maps. Our condition is slightly weaker than that proposed
in \cite{JosJoy02}, but it admits a simple proof and captures the mechanism of chaotic synchronization.
Specifically, it shows that diffusive coupling counteracts intrinsic instability of chaotic system
to produce stable spatially coherent solutions. The synchronizing effect of the coupling is more
pronounced on graphs with localized positive eigenvalues. To this end, in the second part of this paper,
we discuss spectral properties of certain symmetric and random graphs in the light of the sufficient
condition for synchronization. Specifically, we review the relevant facts about the eigenvalues
of Cayley graphs \cite{Biggs} and contrast their properties to those of quasirandom and power-law
graphs \cite{ChuGra88, BarAlb99} and expanders \cite{HooLin06}.
These results highlight the advantages of random connectivity for chaotic synchronization.

\section{Stability analysis}
\setcounter{equation}{0}
Let $\Gamma=(V, E)$ be an undirected
graph on $n$ nodes. The set  of nodes is denoted by $V=[n]:=\{1,2,\dots,n\}$. The edge set $E$ contains
(unordered) pairs of adjacent nodes from $V$.
Unless stated otherwise, all graphs in this paper are assumed to be simple meaning that they do not
contain  loops ($ ii\notin E \;\forall i\in V$) and multiple edges. 
Denote the neighborhood of $i\in V$ by 
$$
N(i)=\{j\in V:\; ij\in E\}.
$$ 
The cardinality of $N(i)$ is called the degree of  node $i\in V$ and denoted by
$d_i=|N(i)|$. If $d_i=d\; \forall i\in [n]$ then $\Gamma$ is called a
$d$-regular graph.

At each node of $\Gamma$ we place a dynamical system
\be\lbl{loc}
x_{k+1}=f(x_k),
\ee
where $f$ is a continuous function from the unit interval $I:=[0,1]$ to itself.
Local dynamical systems at adjacent nodes interact with each other via diffusive coupling. Thus, we have 
the following coupled system
\be\lbl{coup}
x_{k+1}^{(i)}=f(x_{k}^{(i)})+ {\epsilon \over d_i} \sum_{ j\in N(i)} \left(f(x_{k}^{(j)}) -f(x_{k}^{(i)})\right),
\; i\in [n], k=0,1,2,\dots,
\ee
where $\epsilon\ge 0$ controls the coupling strength.

If $\Gamma$ is a lattice, (\ref{coup}) is called a coupled map lattice.  In this paper,
$\Gamma$ can be an arbitrary undirected connected  graph.
For convenience, we rewrite (\ref{coup}) in the vector form
\be\lbl{vec}
\vx_{k+1}=K_\epsilon \vf(\vx_k), \quad K_\epsilon=I_n-\epsilon L,
\ee
where $\vx=(x^{(1)}, x^{(2)},\dots, x^{(n)})$ and
$\vf(\vx_k)=(f(x^{(1)}), f(x^{(2)}),\dots, f(x^{(n)})).$
$L$ stands for the normalized graph Laplacian of $\Gamma$:
\be\lbl{Lap}
L=I_n-D^{-1}A,
\ee
where $I_n$ is the $n\times n$ identity matrix, $D=\operatorname{diag}(d_1,d_2,\dots,d_n)$ is the degree matrix, 
and $A$ is the adjacency matrix of $\Gamma:$
\be\lbl{adj}
(A)_{ij}=\left\{ \begin{array} {ll} 1, & ij\in E, \\ 0, &\mbox{otherwise}.
\end{array}\right.
\ee
Thus,
$$
(L)_{ij}=\left\{ \begin{array} {ll} 1, &i=j,\\ -d_i^{-1}, & ij\in E, \\ 0, &\mbox{otherwise}.
\end{array}\right.
$$
In general, $L$ is not a symmetric matrix. However, it is similar to a symmetric matrix
$$
\tilde L=D^{1/2}L D^{-1/2}=I_n-D^{-1/2}A D^{-1/2}.
$$
Thus, the EVs of $L$ are real and nonnegative by the Gershgorin's
Theorem \cite{HornJohn-Matrix}:
\be\lbl{EVsL}
0\le \lambda_1 \le\lambda_2\le\dots\le\lambda_n.
\ee
Furthermore, the row sums of $L$ are equal to $0$. Thus,  $\lambda_1=0$ and 
\be\lbl{diagonal}
\mathcal{D}=\operatorname{span}\{\1_n\}, \;\mbox{where}\; \1_n:=(1,1,\dots,1)^\t\in\R^n,
\ee
is the corresponding eigensubspace. If $\Gamma$ is connected then $0$ is a simple
EV of $L$ \cite{Fied73}. We will need the following properties of
pseudosimilar transformations (cf.~\cite{Med12}).

\begin{lem}\lbl{lem.intertwin} Let $L$ be the  $n\times n$ Laplacian matrix of a 
connected graph 
$\Gamma$, whose EVs are listed in (\ref{EVsL}).
Suppose $S\in\R^{(n-1)\times n}$ is such that $\operatorname{ker}(S)=\mathcal{D}$ and let
$S^+$ denote the Moore-Penrose pseudoinverse of $S$ \cite{HornJohn-Matrix}.

Then $\hat L=SLS^+$ is  a unique solution of the matrix equation
\be\lbl{matrixEqn}
 \hat LS=SL.
\ee
The EVs of $\hat L$  counting  multiplicities are 
\be\lbl{specD}
\lambda_2, \lambda_3, \dots, \lambda_{n}.
\ee
The eigenspaces of $L$ corresponding to $\lambda_i,~i=2,3,\dots,n,$ 
are mapped isomorphically to the corresponding  eigenspaces of 
$\hat L$ by $S$.
\end{lem}
\pf\; The statement of the lemma follows from Lemma~2.4 of \cite{Med12}.\\

In the remainder of this paper, we use
\be\lbl{defS}
S=
\left(\begin{array}{cccccc}
-1 & 1 & 0& \dots &0& 0 \\
0 & -1& 1 & \dots& 0& 0\\
\dots&\dots&\dots&\dots&\dots&\dots\\
0 &0 & 0& \dots &-1 & 1
\end{array}
\right) \in\R^{(n-1)\times n}.
\ee
Note that $S$ has full row rank and $\operatorname{ker}(S)=\mathcal{D}$. For a given
$\mathbf{x}\in\R^n$, one can use $|S\mathbf{x}|$ to estimate the
distance from $\mathbf{x}$ to 
$\mathcal{D}$. Specifically, for the projection of $\mathbf{x}$ to 
$\mathcal{D}$, $P_{\mathcal{D}},$ we have
\be\lbl{distance}
|P_{\mathcal{D}} \mathbf{x}|=|S^+S\mathbf{x}|\le \| S^+\| |S\mathbf{x}|.
\ee
Here and below, $\|\cdot\|$ stands for the operator norm \cite{HornJohn-Matrix}.

$\mathcal{D}$ is an invariant subspace of (\ref{vec}). Trajectories from this subspace 
correspond to the space homogeneous solutions of (\ref{vec}). Thus, we refer to
$\mathcal{D}$ as a synchronous subspace.  We are interested in finding conditions on 
$\epsilon$, which guarantee (asymptotic) stability of $\mathcal{D}$.
In the analysis below, we follow the approach developed for studying synchronization in 
systems with continuous time \cite{Med11,Med12,MZ12}. 


\begin{thm}\lbl{thm.sync}
Let $f: I\to I$ be a twice continuously differentiable function
and $\Gamma=(V,E)$
be a connected graph on $n$ nodes. Suppose 
\be\lbl{slope}
F=\max_{x\in I} |f(x)|>1,
\ee
and  the EVs of the graph Laplacian
of $\Gamma$ satisfy
\be\lbl{ratio}
{\lambda_{max}\over\lambda_{min}}<1+F^{-1},
\ee
where $\lambda_{min}$ and $\lambda_{max}$ denote the smallest and
largest  positive EVs of the normalized graph Laplacian $L,$ respectively.

Let $\mathbf{x}_k, k=0,1,2,\dots$ denote a trajectory of (\ref{coup}) with 
\be\lbl{domain}
\epsilon\in \left(\lambda_2^{-1}(1-F^{-1}),
  \lambda_n^{-1}(1+F^{-1})\right).
\ee

Then there exists $\delta>0$ such that $|P_{\mathcal{D}} \mathbf{x}_k|\to 0$ as $k\to\infty$
provided $|P_{\mathcal{D}} \mathbf{x}_0|<\delta$. 
\end{thm} 

For the proof this theorem, we will need the following auxiliary lemma, which we state 
first.
\begin{lem}\lbl{lem.nonlin}
Let $\{u_k\}_{k=0}^\infty$ be a sequence of nonnegative numbers such that
\be\lbl{growth}
u_k\le \mu^ku_{k-1}+A \sum_{j=0}^{k-1}\mu^j u_{k-1-j}^2,\; k=1,2,3,\dots,
\ee
for some $0<\mu<1$ and $A>0$. Then there exist positive numbers $\eta_0$ and $\eta$ such that
\be\lbl{geometric}
u_k\le \eta\mu^k, \; k=0,1,2,\dots,
\ee
provided $0\le u_0 \le \eta_0$.
\end{lem}

\pf (Theorem~\ref{thm.sync}) Let $\mathbf{x}_k, k=0,1,2,\dots,$ denote a trajectory of (\ref{vec}).
By multiplying both sides of (\ref{vec}) by $S$, we have 
\be\lbl{yk}
\mathbf{y}_k=SK_\epsilon \vf(\vx_k)=\hat K_\epsilon S\vf(\vx_k),\;
\hat K_\epsilon=I_{n-1}-\epsilon \hat L,
\ee
where $\vy_k=S\vx_k$. Decompose
\be\lbl{decomp}
\vx_k=c_k\1_n + \vv_k,\; c_k=n^{-1}\1_n^\t\vx_k, \; \vv_k=S^+S\vx_k\in\mathcal{D}^\perp,
\ee
where we are using the fact that $S^+S$ provides the projection on 
$\operatorname{ker}(S)^\perp=\mathcal{D}^\perp$. Using Taylor's formula,
we have
\be\lbl{rewrite-rhs}
  S\vf(\vx_k)=S\left\{f(c_k)\1+ f^\prime(c_k) \vv_k+\tilde \vg_k(\vv_k)\right\}=
f^\prime(c_k) \vy_k+ \vg_k(\vy_k),
\ee
where $\vy_k=S\vx_k=S\vv_k$, $\vg_k(\vy_k)=S\tilde \vg_k(S^+\vy_k)$, 
\be\lbl{quad}
|\vg_k(\vy)| \le C_1 |\vy|^2,
\ee
and $C_1$ is a positive constant independent of $k\in\N\cup \{0\}$.
By plugging (\ref{rewrite-rhs}) in (\ref{yk}), we have
\be\lbl{mapy}
\vy_{k+1}=f_k \vy_k + \vg_k,
\ee 
where $f_k:=f^\prime(c_k)$ and $\vg_k:=\vg_k(\vy_k)$.

By iterating (\ref{mapy}), we obtain
\be\lbl{iterate}
\vy_k=\left[ \prod_{j=0}^{k-1} (f_j \hat K_\epsilon)\right]  \vy_0 +
\sum_{j=0}^{k-1}  \left[ \prod_{l=j+1}^{k-1} (f_l \hat K_\epsilon) \right] \vg_j,
\ee
where the product over
an empty index set  is assumed to be equal to the identity matrix.

Condition (\ref{ratio}) implies that $\left(\lambda_2^{-1}(1-F^{-1}), \lambda_n^{-1}(1+F^{-1})\right)$
is nonempty.  By Lemma~\ref{lem.intertwin}, the EVs of $\hat L$ lie between $\lambda_2$
and $\lambda_n$. Thus, 
$\mu:=F\|I_{n-1}-\epsilon \hat L\|<1$,
whenever
$$
\epsilon\in \left(\lambda_2^{-1}(1-F^{-1}),
  \lambda_n^{-1}(1+F^{-1})\right).
$$
Thus, 
\be\lbl{contract}
|f_j|\|K_\epsilon\|\le \mu<1, \; j\in \N\cup \{0\}.
\ee

Using (\ref{quad}) and (\ref{contract}), from (\ref{iterate}) we obtain
\be\lbl{geom}
|\vy_k | \le \mu^k |\vy_0| + C_1 \sum_{j=0}^{k-1} \mu^j |\vy_{k-1-j}|^2,\;
k\in\N.
\ee
Using Lemma~\ref{lem.nonlin}, from (\ref{geom}) we obtain
$$
|\vy_k|\le C_2 \mu^k, \; k\in \N\cup \{0\},
$$
provided $\vy_0$ is sufficiently small.\\
$\qed$

\pf (Lemma~\ref{lem.nonlin})

We use induction to prove (\ref{geometric}). Let  $\eta>0$ be fixed. Denote
\be\lbl{def.xi}
f(\xi)=\xi\exp\{ -A\xi/\mu(1-\mu)\}
\ee
and set $\eta_0:=f(\eta)\le \eta$.
Clearly, (\ref{geometric}) is true for $k=0$ and $0\le u_0\le\eta$.

Suppose (\ref{geometric}) holds for $0\le k<p$ for some $p\in\N$. Below we show  
that this implies that it also holds for $k=p$. To this end, let
$$
z_k=u_k\mu^{-k},\; k=0,1,2,\dots.
$$
Using (\ref{growth}), we have 
\begin{eqnarray}\nonumber
z_p &\le& u_0+A \sum^{p-1}_{j=0} \mu^{-1} u_{p-1-j} z_{p-1-j}\\
\nonumber
&\le&  u_0+A\eta \sum^{p-1}_{j=0} \mu^{p-j-2} z_{p-1-j}\\
\lbl{step}
&=& u_0+A\eta \sum_{j=1}^p \mu^{j-2}z_{j-1},
\end{eqnarray}
where we used the induction hypothesis in the second inequality.

Using the Gronwall's inequality (see Lemma~\ref{lem.Gron} below), from (\ref{step})
we obtain
\be\lbl{use-Gron}
z_p\le u_0\exp\{A\eta\sum_{j=1}^p \mu^{j-2}\} 
\le u_0 \exp\{A\eta/ \mu(1-\mu)\}= {u_0\eta\over f(\eta)}\le \eta,
\ee
where we used $0\le u_0\le\eta\le f(\eta)$ in the last inequality.
Thus, 
$$
u_p\le \eta\mu^p.
$$
The statement of the lemma follows by induction.\\
$\qed$

The proof of following discrete counterpart of the Gronwall's inequality can
be found in \cite{KocPal10}.

\begin{lem}\lbl{lem.Gron}
Let $\{z_k\}_{k=0}^\infty$ and $\{\mu_k\}_{k=1}^\infty$ be two nonnegative
sequences such that
\be\lbl{growth}
z_k\le B+\sum_{j=1}^{k}\mu_j z_{j-1}, \; k\in [p],
\ee
for $B\ge 0$ and $p\in\N$. Then 
$$
z_k\le B\exp\left\{\sum_{j=1}^k \mu_j\right\}, \; k\in [p].
$$
\end{lem}

\section{Connectivity}\lbl{sec.connectivity}
\setcounter{equation}{0}
In this section, we examine the role of the network topology for 
chaotic synchronization in the light of Theorem~\ref{thm.sync}. 
The sufficient condition for synchronization (\ref{ratio})  favors graphs
whose nonzero EVs are localized. Thus, we review several families of graphs
$\Gamma_n=(V(\Gamma_n), E(\Gamma_n)),$ $n\in\N,$  which possess this property 
even when $|V(\Gamma_n)|$ grows without bound as $n\to\infty$.
These families include   Erd{\"o}s-R{\'e}nyi (ER) random,  quasirandom,
 and power-law graphs \cite{ChuGra88, KriSud06, BarAlb99}, and expanders\cite{HooLin06}. 
We also review  spectral properties of Cayley graphs. 
Because Theorem~\ref{thm.sync} provides only a sufficient condition for synchronization,
we tested how well (\ref{domain}) captures the domain of
synchronization numerically.
We found a good agreement between numerical results and  the analytical estimates
of the synchronization domain.
Numerics also confirm 
that localization of the graph EVs is critical for synchronization of      
large networks.
 
\subsection{Cayley graphs}\lbl{sec.Cayley}
We start our review of network connectivity with Cayley graphs. These
are highly symmetric graphs defined on groups.

\begin{df}\lbl{df.Cayley}
Suppose $G$ is a finite additive group and $S\subset G$ is a symmetric subset, i.e., $S=-S$.
Let $\Gamma=\langle V, E\rangle$ be a graph defined as follows
$V=G$ and for any $a,b\in G,$ $ab\in E$ if $b-a\in S$.
$\Gamma$  is called a Cayley graph and is denoted $\Cay(\Gamma,S)$.
\end{df}  

\begin{figure}
\begin{center}
{\bf a}\hspace{0.1 cm}\includegraphics[height=1.8in,width=2.0in]{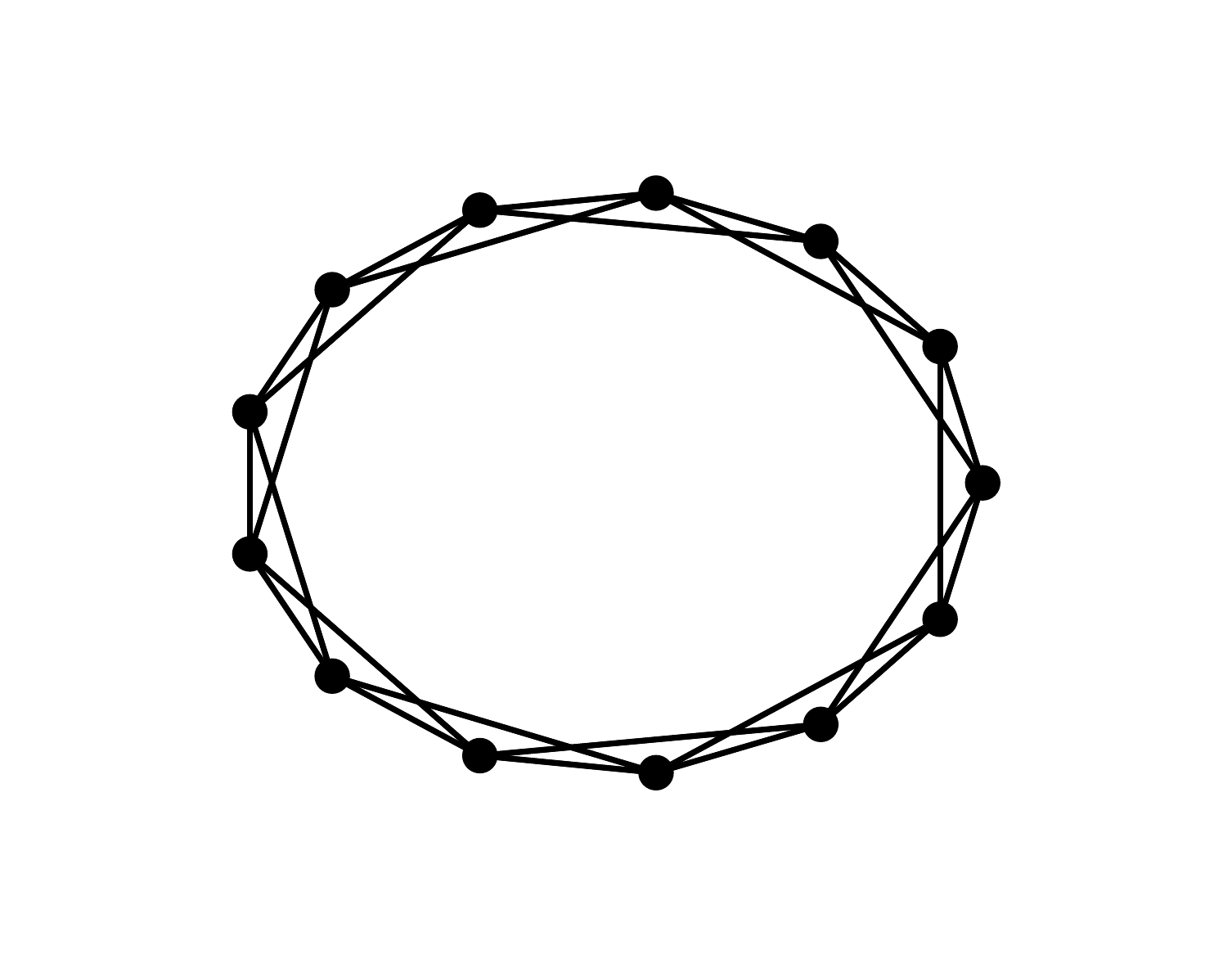}
{\bf b}\hspace{0.1 cm}\includegraphics[height=1.8in,width=2.0in]{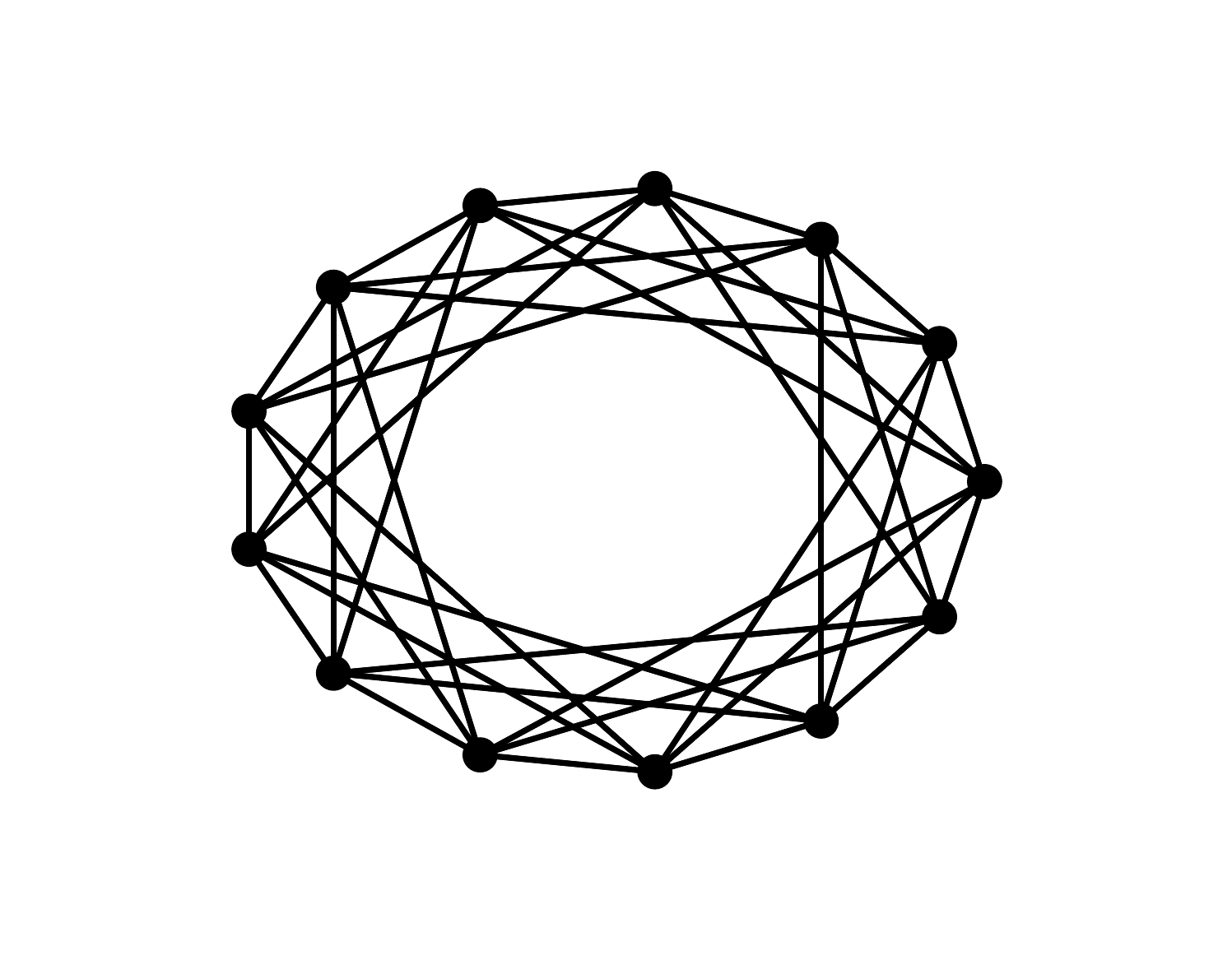}
{\bf c}\hspace{0.1 cm}\includegraphics[height=1.8in,width=2.0in]{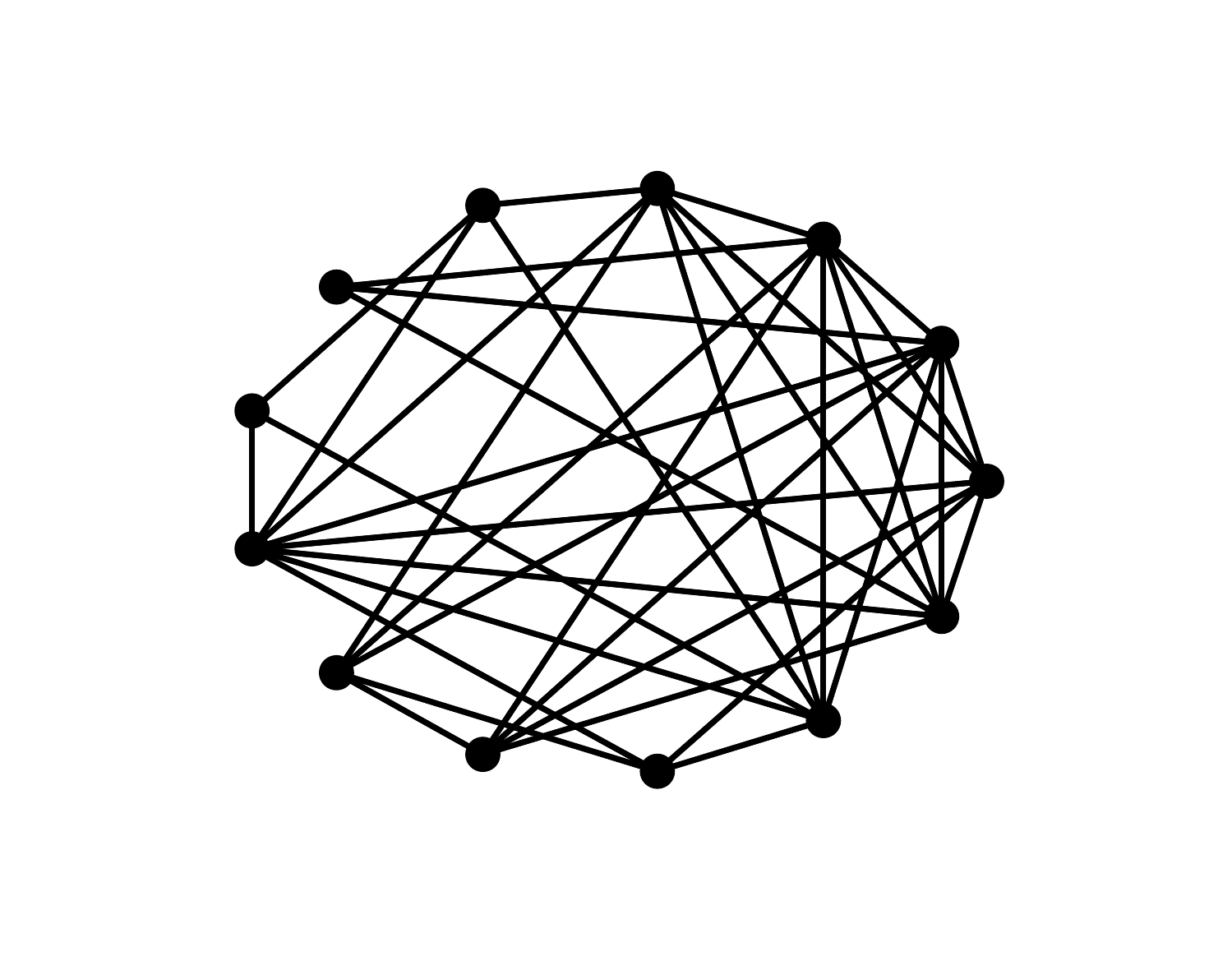}
\end{center}
\caption{ {\bf a})  A Cayley graph based on a ball, {\bf b}) a Paley graph, and  {\bf c}) an ER graph. 
}
\lbl{f.1}
\end{figure}
The following  Cayley graphs on  the (additive) cyclic group
$\Z_n=\Z/n\Z, n\in\N,$ model nearest-neighbor connectivity in (\ref{coup}).  
\begin{ex}\lbl{ex.ball}
Let $n\ge 3, r\le\lfloor n/2\rfloor$ and $B(r)=\{\pm 1, \pm 2, \dots, \pm r \}.$
$\Gamma =\Cay(\Z_n, B(r))$ is called a Cayley graph based on the ball $B(r)$.
\end{ex}
Cayley graph $\Gamma=\Cay(\Z_n,B(1))$ corresponds to the standard nearest-neighbor 
interactions  in the coupled system  (\ref{coup}). By varying  $r$ in $\Cay(\Z_n, B(r)),$  
one can control the range of coupling (see Fig.\ref{f.1}a). The analysis of the coupled 
system (\ref{coup}) on $\Gamma=\Cay(\Z_n,B(r))$ below shows that for fixed $r$ and $n\gg1$
they do not meet synchronization condition (\ref{ratio}). This will be contrasted with the 
analysis of networks on random and quasirandom graphs, which will be
discussed in \S\ref{sec.rand}.

The EVs of Cayley graphs on $\Z_n$ can be computed explicitly in terms of the one-dimensional 
representations of $\Z_n:$
\be\lbl{ex}
\mathbf{e}_x=(\mathbf{e}_x(0), \mathbf{e}_x(1), \dots,
\mathbf{e}_x(n-1))^\t, \; x\in\Z_n,
\ee
where
\be\lbl{character}
e_x(y)=\exp\left\{ {2\pi\iu xy\over n}\right\}, \; y\in\Z_n.
\ee

For the EVs of Cayley graphs on a cyclic group, the following lemma is well-known \cite{KriSud06}.

\begin{lem}\lbl{lem.EV-cyc}
The EVs of the normalized graph Laplacian of $\Gamma=\Cay(\Z_n, S)$ are given by
\be\lbl{EV-C}
\lambda_x= 1-|S|^{-1} \sum_{y\in S} \mathbf{e}_x(s)=1-|S|^{-1}
\sum_{y\in S}\cos\left({2\pi xy\over n}\right), \; x\in\Z_n.
\ee
The corresponding eigenvectors are  $\mathbf{e}_x, \; x\in\Z_n.$
\end{lem}
\pf
For any $x,y\in\Z_n,$ we have
$$
(A\mathbf{e}_x)(y)=\sum_{s\in S}\mathbf{e}_x(y+s)=\left(\sum_{s\in S}\mathbf{e}_x(s)\right) \mathbf{e}_x(y).
$$
Thus,
\be\lbl{EV-AC}
A\mathbf{e}_x=\mu_x \mathbf{e}_x, \quad \mu_x=\sum_{s\in S}\mathbf{e}_x(s).
\ee
Since characters $\mathbf{e}_x, x\in\Z_n,$ are mutually orthogonal, (\ref{EV-AC}) 
gives the full spectrum of $A$. The statement of the lemma follows from  (\ref{EV-AC}) and $L=|S|I_n-A$.\\
$\qed$

\begin{cor}\lbl{cr.zero} The spectrum of  $\Gamma=\Cay(\Z_n, S)$ contains a simple zero EV
$
\lambda_0=0
$
corresponding to the constant eigenvector $\mathbf{1_n}=(1,1, \dots,1)^\t$.
\end{cor}

With the explicit formula for the EVs of the Cayley graph $\Cay(\Z_n, S)$ 
in hand (cf.~(\ref{EV-C})), we study synchronization in the coupled system (\ref{coup}) on symmetric graphs.
We start with the nearest-neighbor coupling corresponding to $\Gamma=\Cay(\Z_n, B(1))$.

\begin{ex}\lbl{ex.B1}
By applying (\ref{EV-C}) to $\Gamma=\Cay(\Z_n, B(1))$, we obtain the following 
formula for the nonzero EVs of $\Gamma$:
\be\lbl{nonzeroEV-B1}
\lambda_x=1-\cos\left({2\pi x\over n}\right),\; x\in [n-1].
\ee
By setting $x=1$ and $x=\lfloor {n\over 2}\rfloor$ in (\ref{nonzeroEV-B1}), we find the smallest 
and the largest positive EVs of $\Gamma$, respectively:
\begin{eqnarray}\lbl{lmin-B1}
\lambda_{min} &=& 1-\cos\left({2\pi \over n}\right)= {2\pi^2\over n^2}+ O(n^{-4}),\\
\lbl{lmax-B1}
\lambda_{max} &=& \left\{ \begin{array}{ll} 2, & \mbox{if}\; n\; \mbox{is even},\\
            1-\cos(\pi(1-n^{-1}))= 2+ O(n^{-2}),  & \mbox{if}\; n\; \mbox{is odd},
\end{array}
\right.
\end{eqnarray}
Thus,
\be\lbl{nearest-ratio}
{\lambda_{max}\over\lambda_{min}}={n^2\over \pi^2} + O(n^{-2}).
\ee
We conclude that the nearest-neighbor family of graphs does not satisfy synchronization
condition (\ref{ratio}) for large $n$.
\end{ex}

Following \cite{JosJoy02}, in our numerical experiments we use
the following measure of coherence. For  $\{\vx_k, k\in\N\},$ a trajectory of (\ref{coup}), 
consider 
\be\lbl{sigma}
(\sigma^2 (x))_k={1\over n} \sum_{j=1}^n (x_k^{(j)}-\bar \vx_k)^2, \quad \bar \vx_k= 
{1\over n}\sum_{j=1}^n x_k^{(j)}, k=0,1,2,\dots.
\ee
We will refer to $\sigma^2 (\vx)$ as the variance of $\vx$. Clearly, synchronization means 
that  $|(\sigma^2 (\vx))_k|$ stays small for $k\gg 1$.
Numerical results for (\ref{coup}) on several nearest-neighbor graphs are shown in Fig.~\ref{f.NN}.
The plots in Fig.~\ref{f.NN} show the range of values of $\sigma^2$ (after removing transients)
for a typical trajectory for the range of coupling strength $\epsilon\in (0,1]$
for the nearest-neighbor graphs of different size. The windows in $\epsilon$ where the range 
of $\sigma^2$ is close to $0$ indicate synchronization. Plot \textbf{a)} for $n=5$ shows
a window of synchrony for an interval in $\epsilon$, which shrinks for $n=6$ (Fig.~\ref{f.NN} \textbf{b}); 
and  already for $n=7$ (Fig.~\ref{f.NN} \textbf{c}) no windows of synchronization were found.
\begin{figure}
\begin{center}
{\bf a}\hspace{0.1 cm}\includegraphics[height=1.8in,width=2.0in]{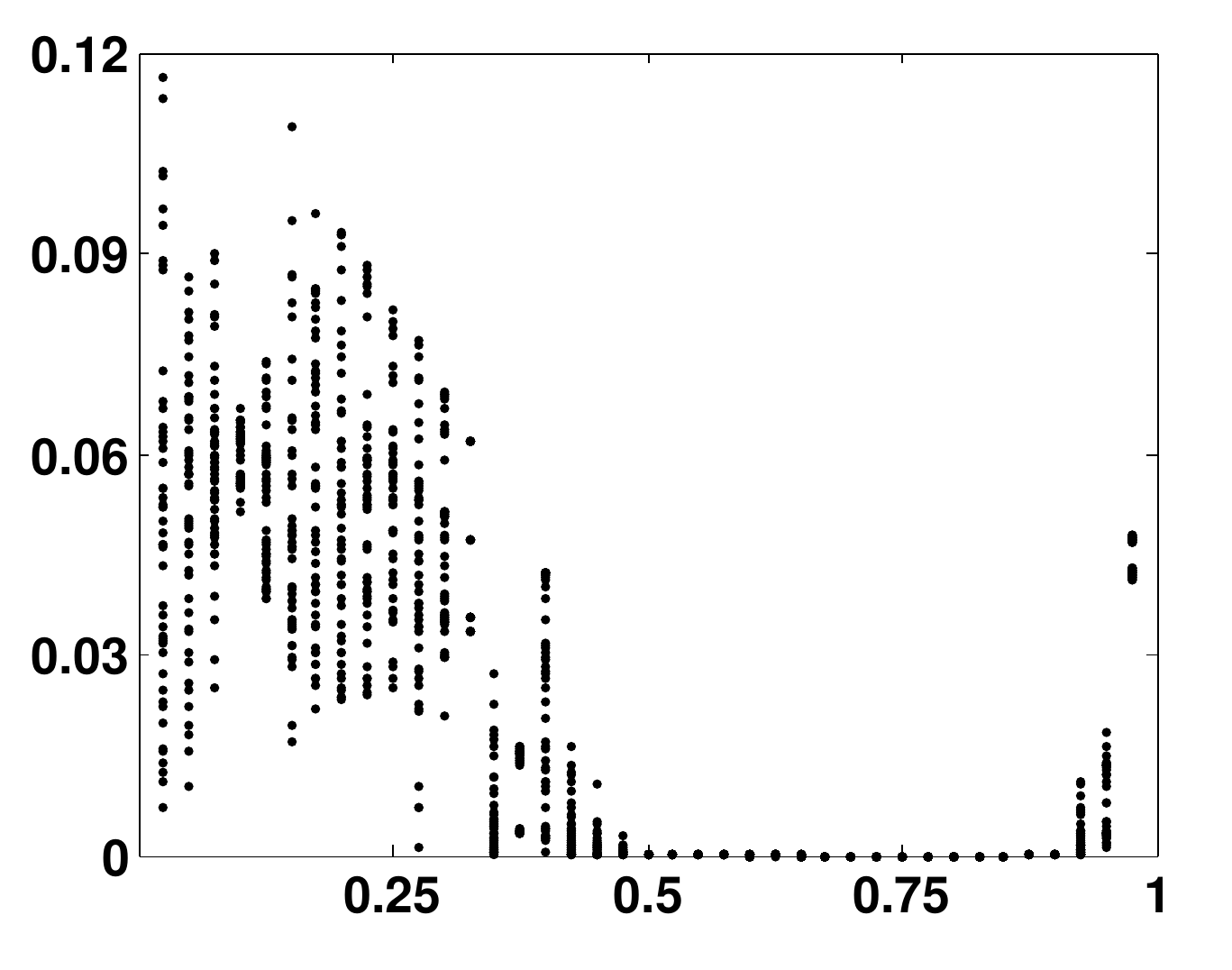}
{\bf b}\hspace{0.1 cm}\includegraphics[height=1.8in,width=2.0in]{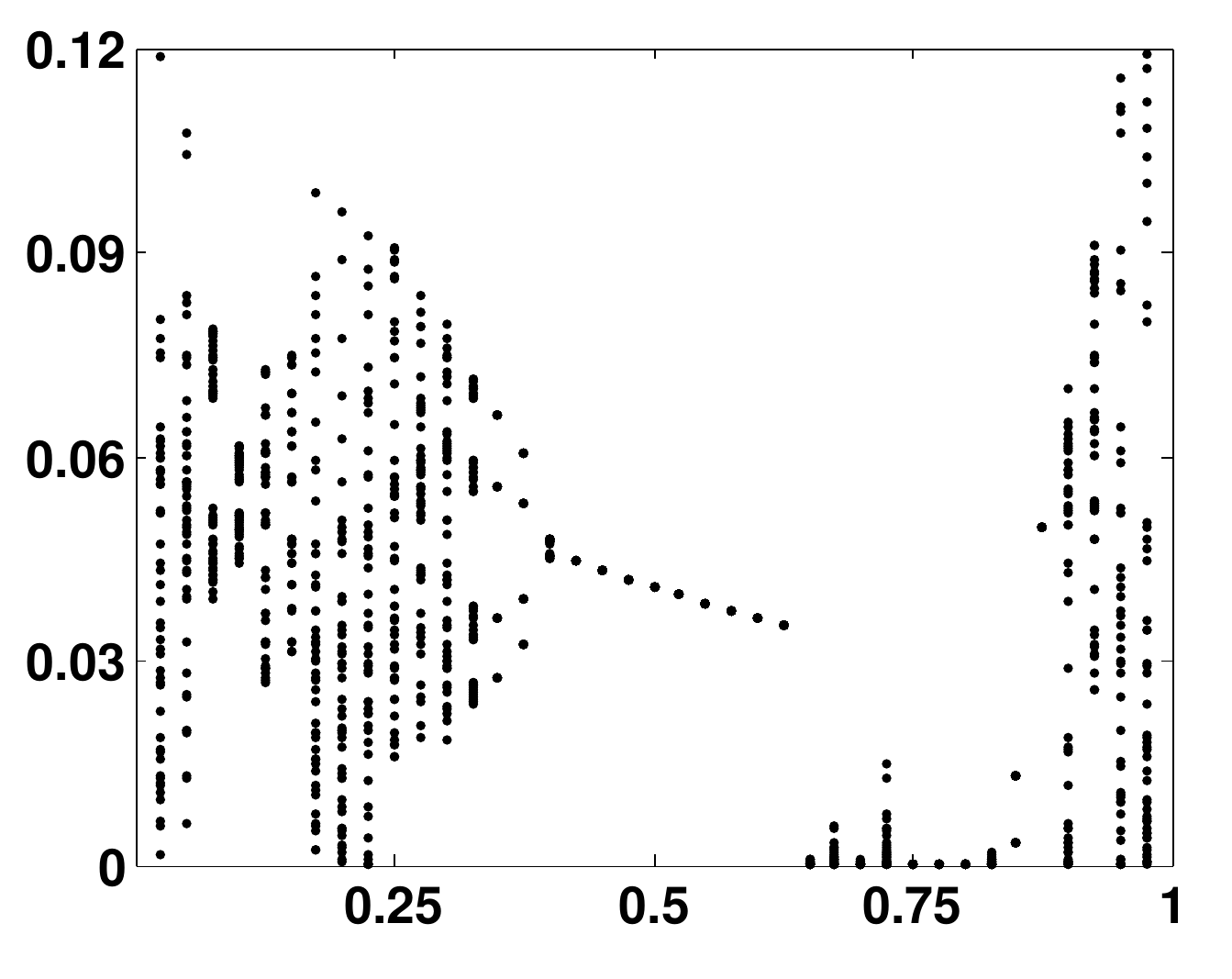}
{\bf c}\hspace{0.1 cm}\includegraphics[height=1.8in,width=2.0in]{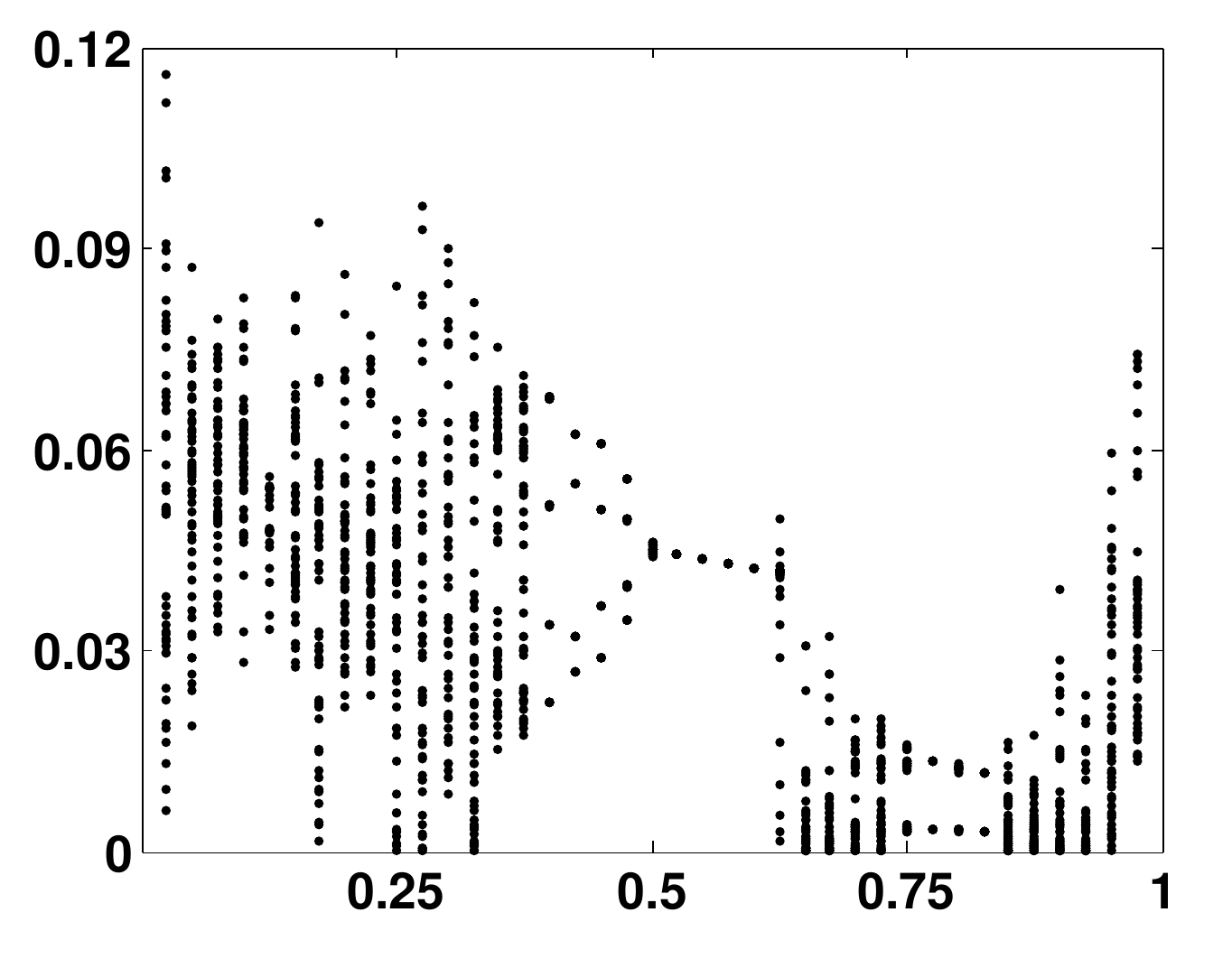}
\end{center}
\caption{ The range of the variance of a typical trajectory of (\ref{coup}) 
on $\Cay(\Z_n, B(1))$ as a function of the coupling strength $\epsilon$:
\textbf{a)} $n=5$, \textbf{b)} $n=6$, and  \textbf{c)} $n=7.$
}
\lbl{f.NN}
\end{figure}

\begin{figure}
\begin{center}
{\bf a}\hspace{0.1 cm}\includegraphics[height=1.8in,width=2.0in]{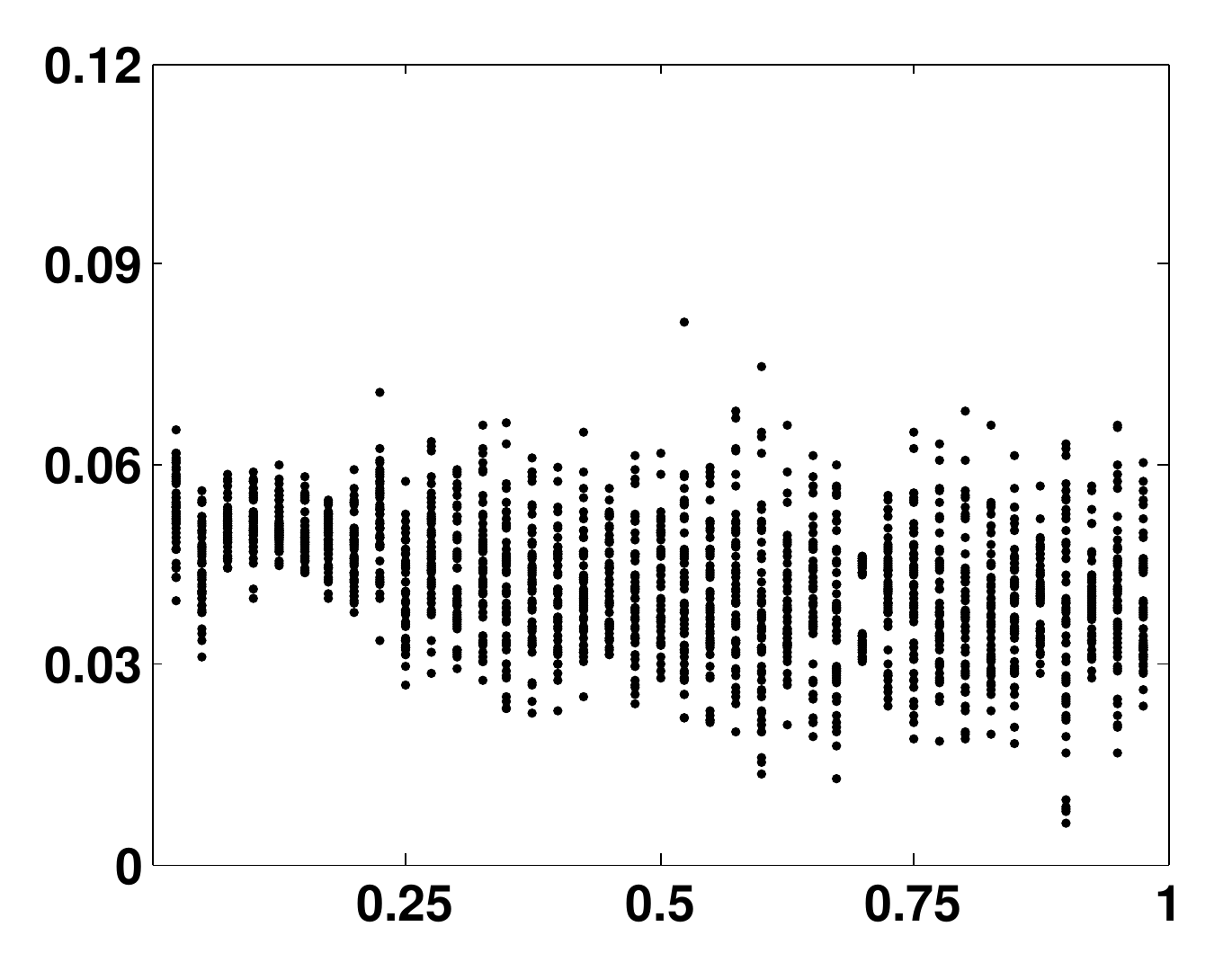}
{\bf b}\hspace{0.1 cm}\includegraphics[height=1.8in,width=2.0in]{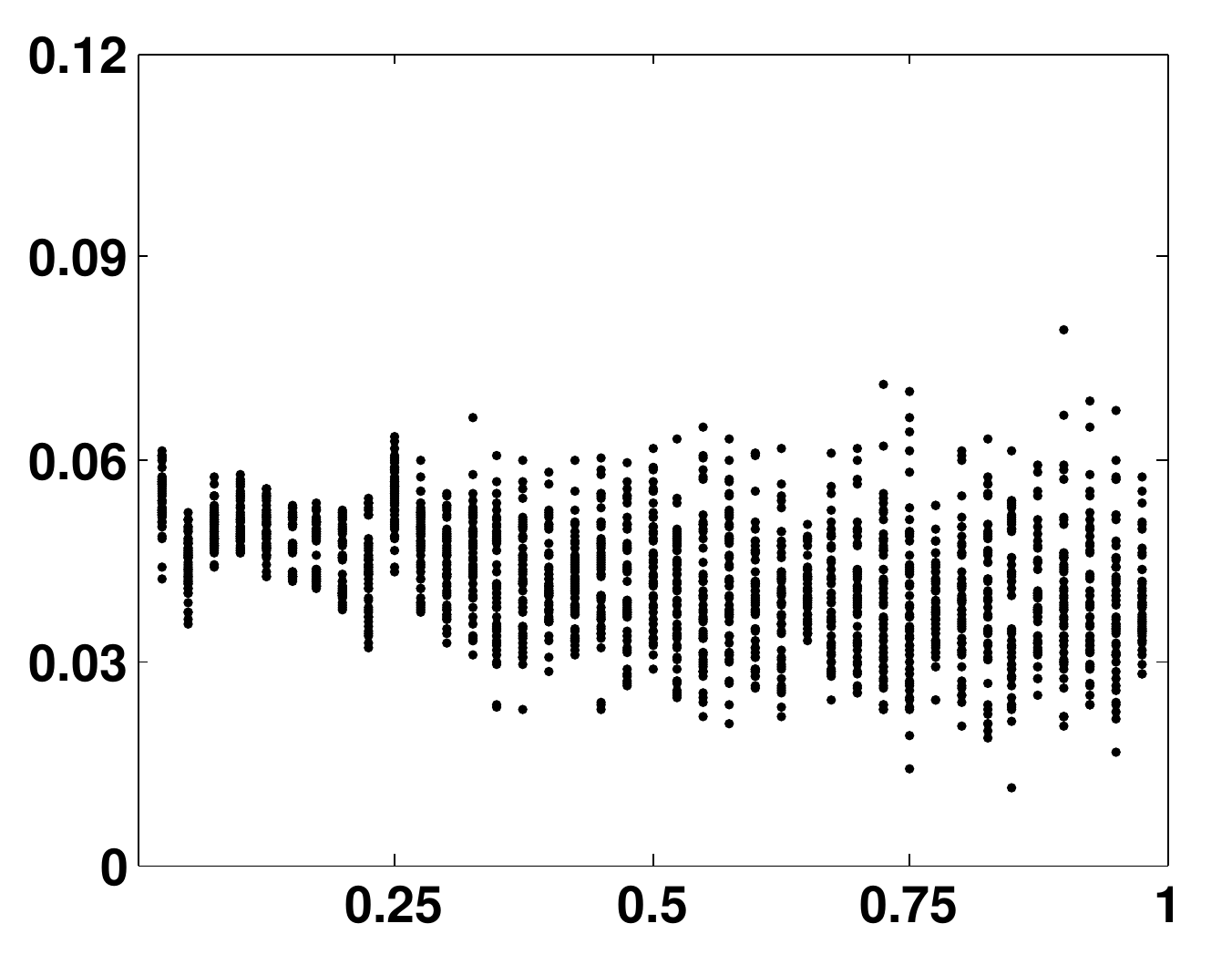}
{\bf c}\hspace{0.1 cm}\includegraphics[height=1.8in,width=2.0in]{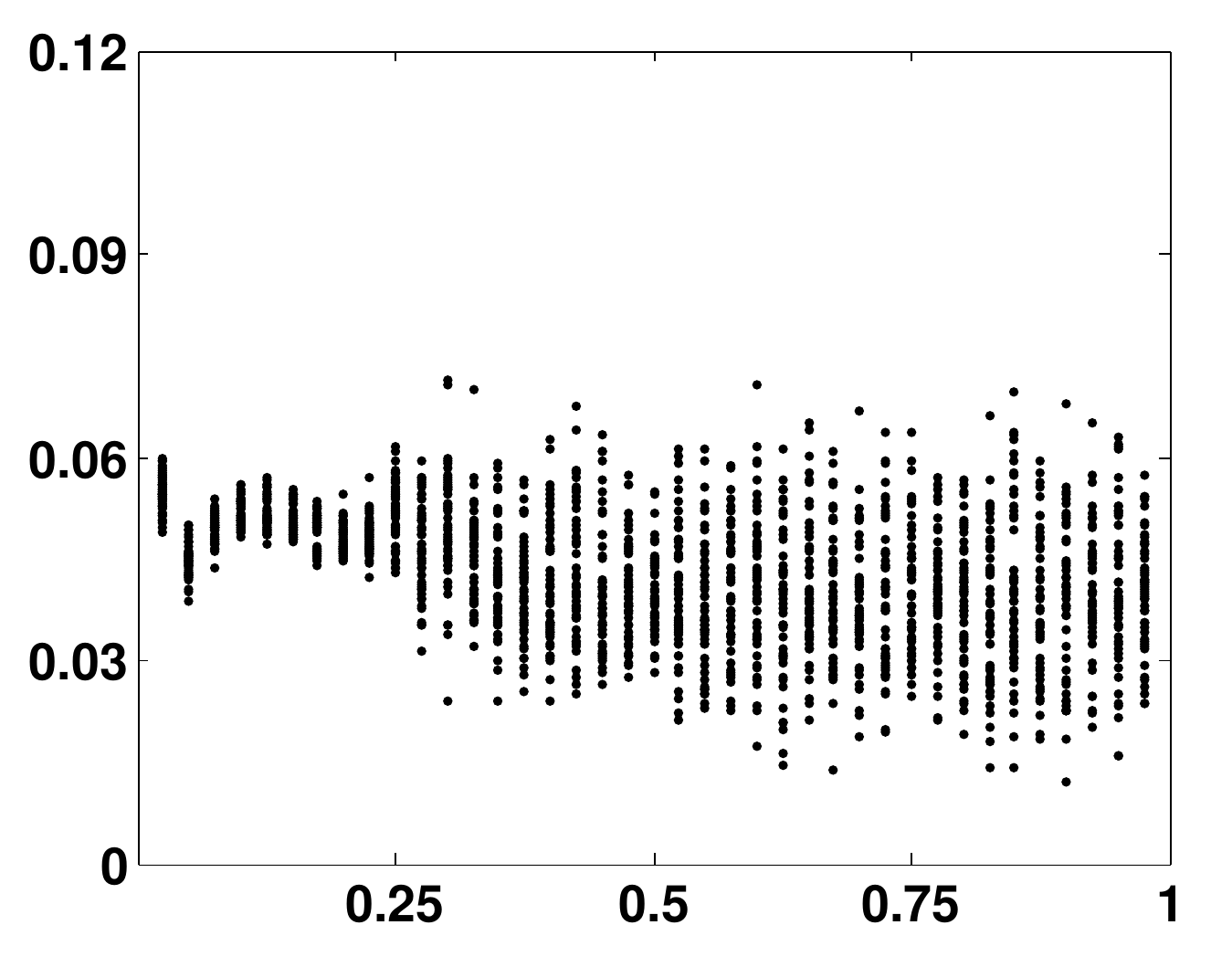}\\
{\bf d}\hspace{0.1 cm}\includegraphics[height=1.8in,width=2.0in]{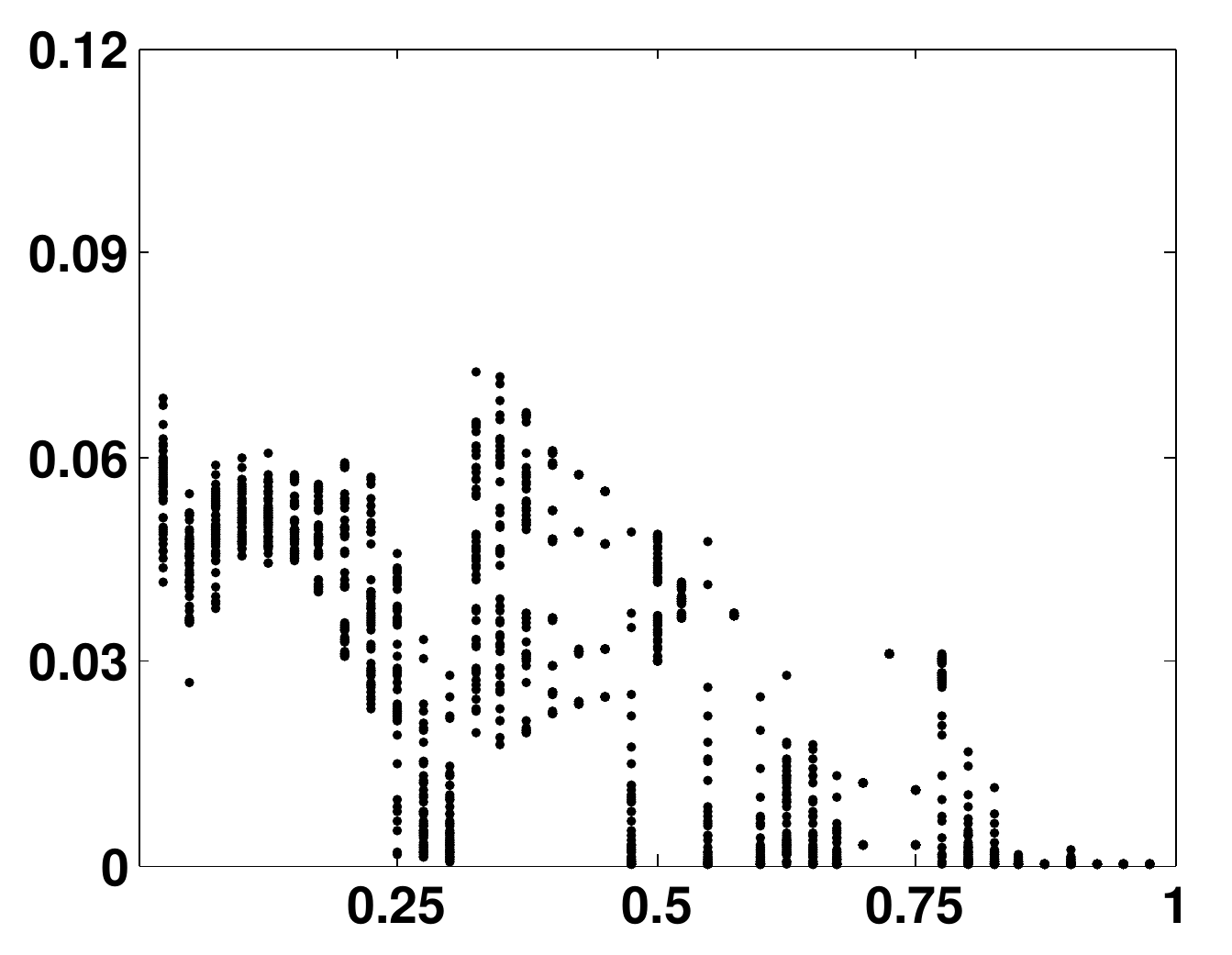}
{\bf e}\hspace{0.1 cm}\includegraphics[height=1.8in,width=2.0in]{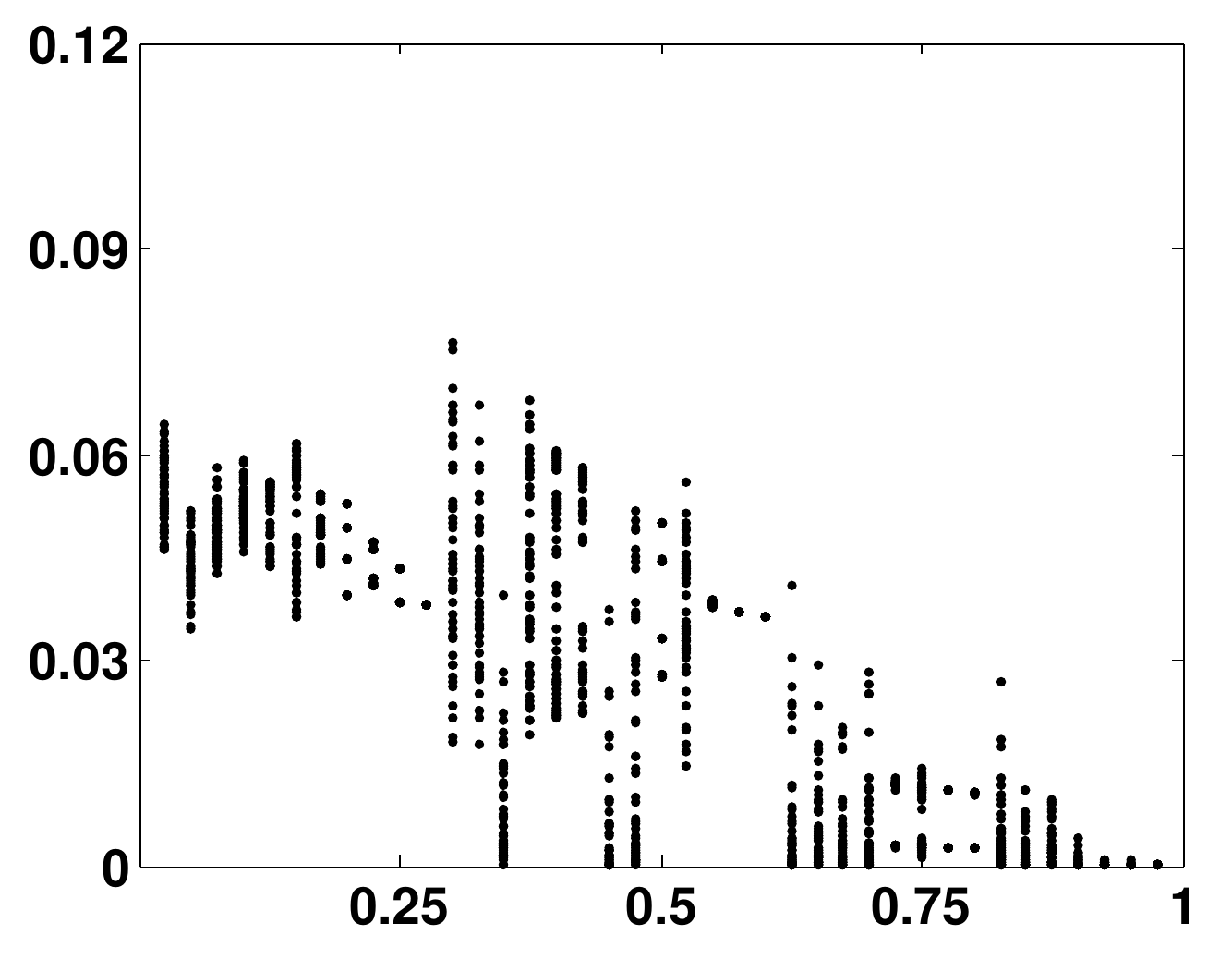}
{\bf f}\hspace{0.1 cm}\includegraphics[height=1.8in,width=2.0in]{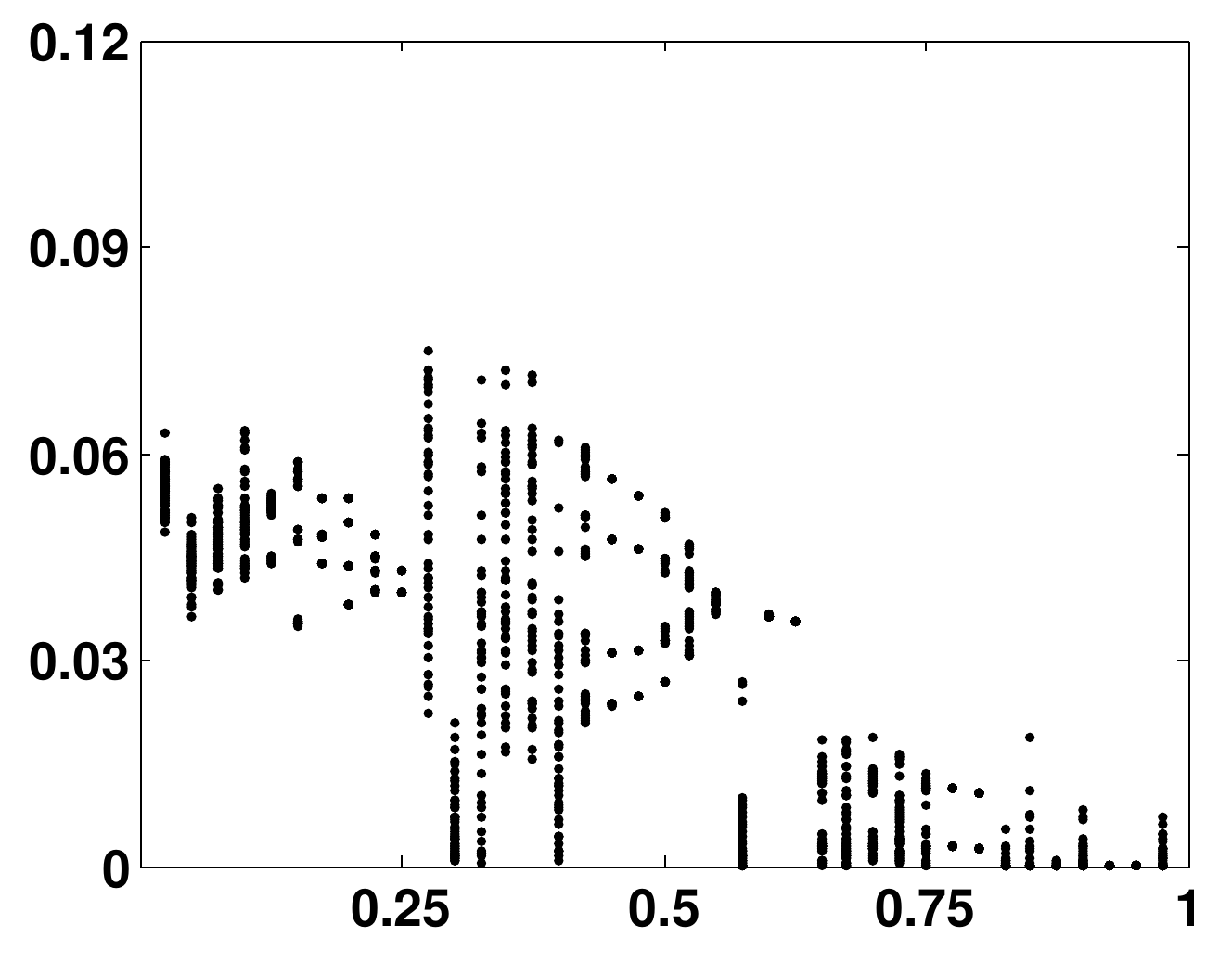}\\
{\bf g}\hspace{0.1 cm}\includegraphics[height=1.8in,width=2.0in]{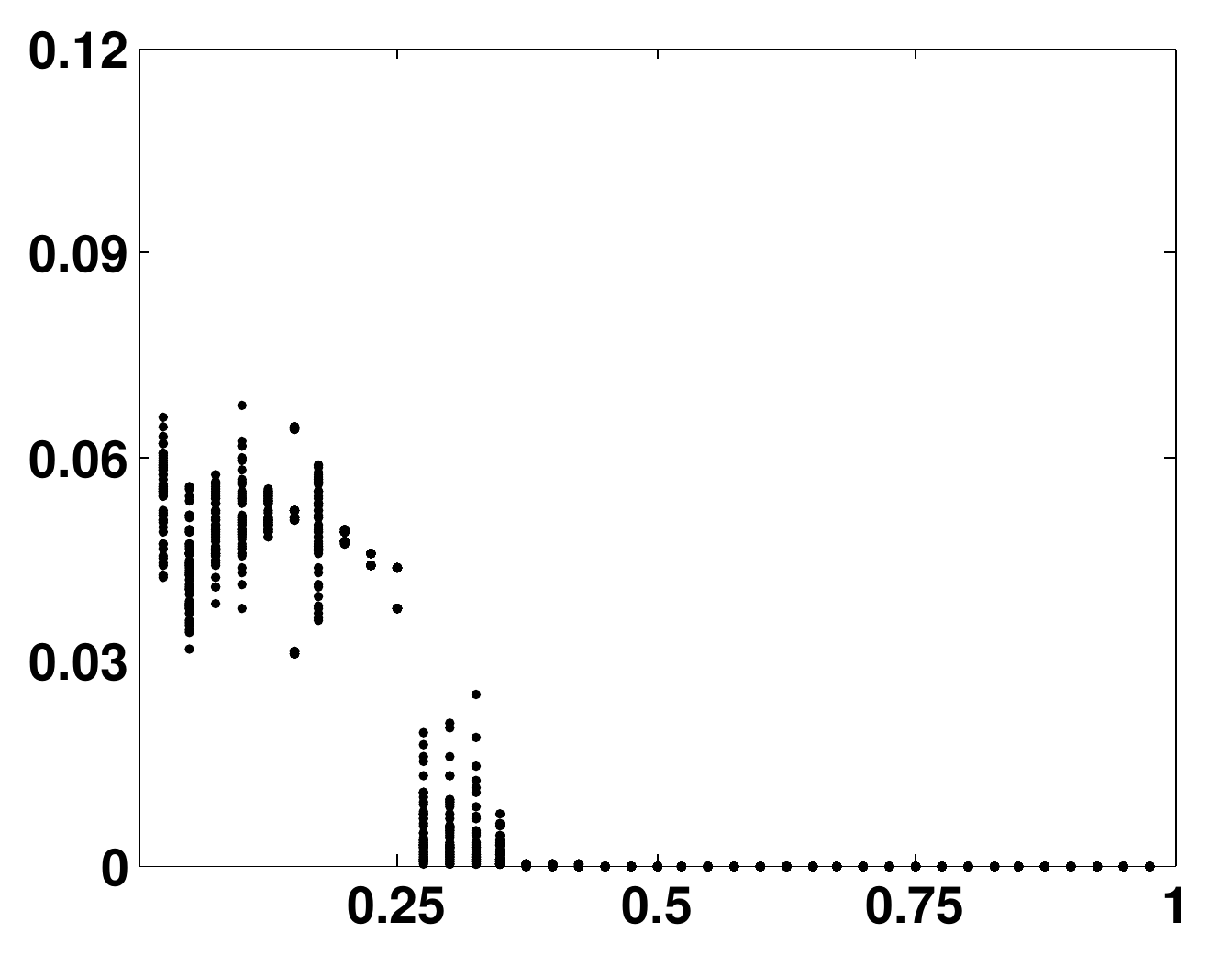}
{\bf h}\hspace{0.1 cm}\includegraphics[height=1.8in,width=2.0in]{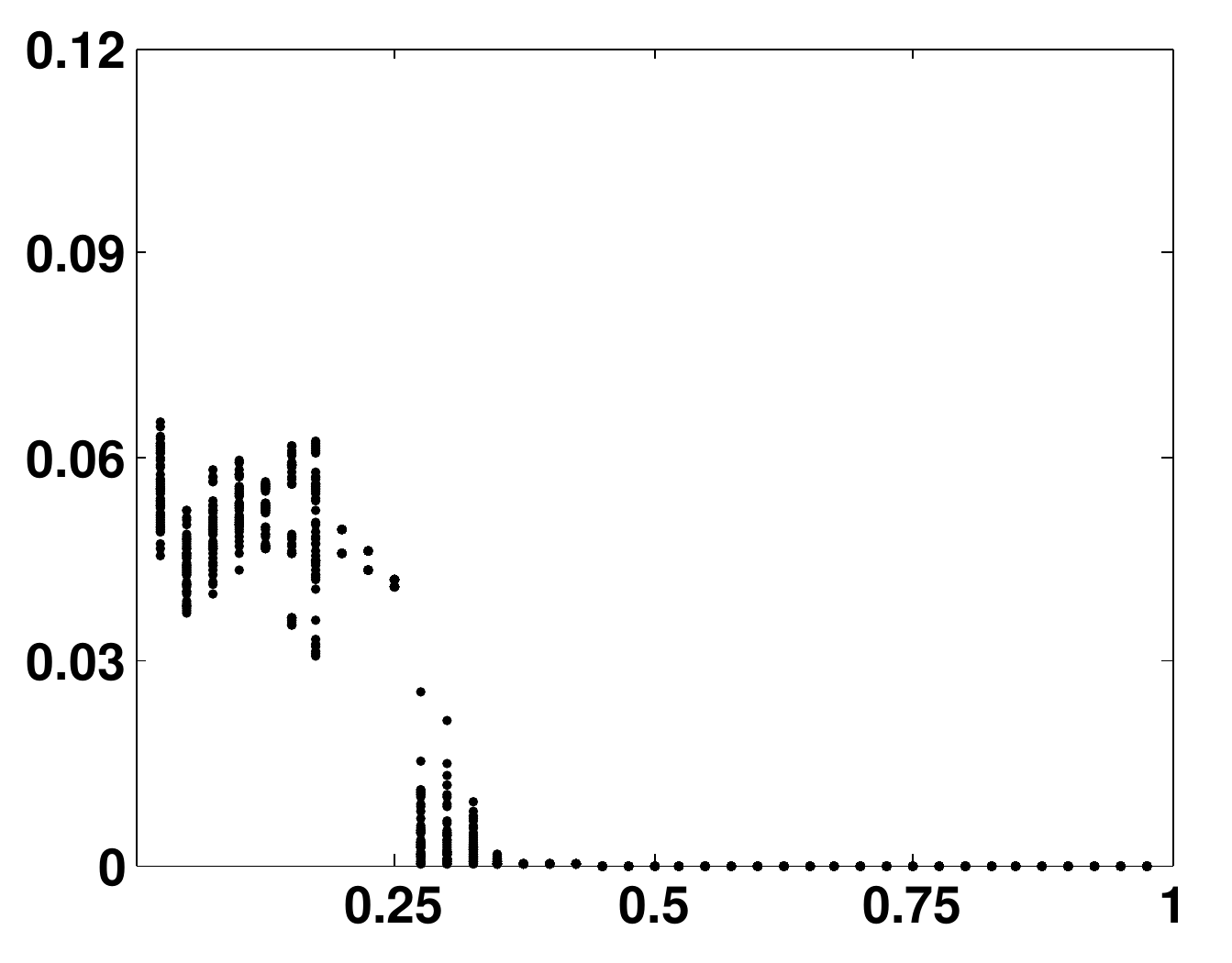}
{\bf i}\hspace{0.1 cm}\includegraphics[height=1.8in,width=2.0in]{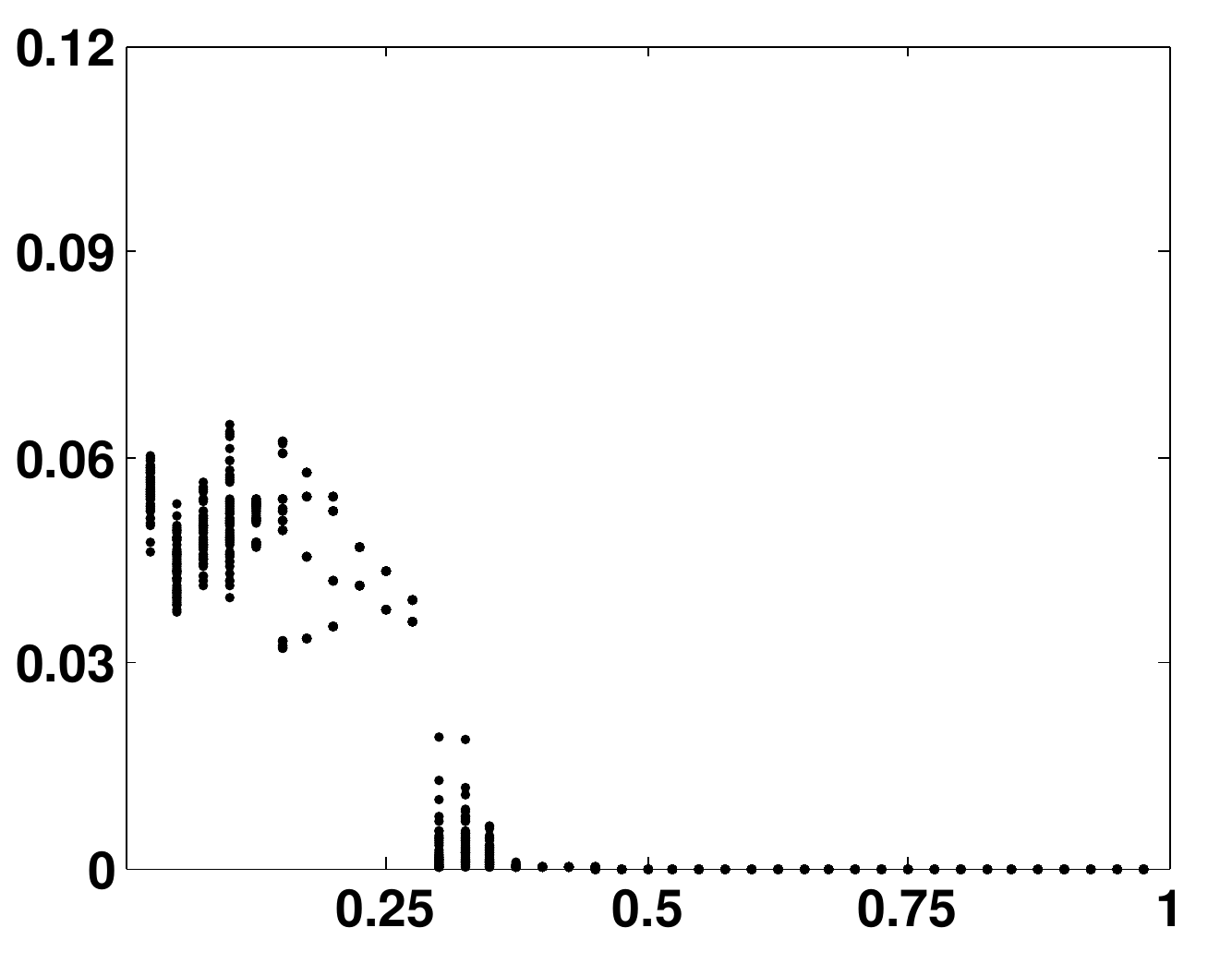}
\end{center}
\caption{ The range of the variance of a typical trajectory of (\ref{coup}) 
on $\Cay(\Z_n, B(r))$ as a function of the coupling strength $\epsilon$:
 {\bf a})  $n=100,$ $r=10$, 
\textbf{b)} $n=200,$ $r=20,$ and  \textbf{c)} $n=400,$ $r=40$, \textbf{d)} $n=100,$ $r=25$, 
\textbf{e)} $n=200,$ $r=50,$ and  \textbf{f)} $n=400,$ $r=100$, \textbf{g)} $n=100,$ $r=49$, 
\textbf{h)} $n=200,$ $r=98,$ and  \textbf{i)} $n=400,$ $r=196$.
}
\lbl{f.Cay}
\end{figure}

We now turn to $\Gamma=\Cay(\Z_n, B(r))$  for  $r\in \N$ and $n>2r$. This is the case of the
$r$-nearest-neighbor coupling. 
\begin{ex}\lbl{ex.Br} For $\Gamma=\Cay(\Z_n, B(r))$, using (\ref{EV-C}),
we have
\begin{eqnarray}\nonumber
\lambda_x& =& 1-{1\over 2r} \left( \sum_{k=-r}^{r} \cos \left({2\pi x k\over n}\right) -1\right)\\
\lbl{EV-Br}
&=& 1-{1\over 2r} \left( {\sin\left({\pi x(2r+1)\over n}\right)\over
    \sin\left({\pi x\over n}\right)} -1 \right), \; x\in [n-1].
\end{eqnarray}
Using the Taylor's formula, after straightforward calculations from (\ref{EV-Br}) we estimate
\be\lbl{lmin-Br}
0<\lambda_{min}\le \lambda_1  = {\pi^2(2r+1)(r+1)\over 3n^2}+ O(n^{-4}).
\ee
To estimate $\lambda_{max}$ we use $x^*=\lfloor {n\over 2r+1}\rfloor$.
For fixed $r\ge 1$ and $n\gg 1$, using Taylor's formula and elementary inequalities, we
estimate
\begin{eqnarray*}
\sin\left({\pi\over 2r+1}\right) &\le &\sin\left({\pi x^*\over n}\right) \le 
\sin\left({\pi\over 2r+1}\right) +O(n^{-1}),\\
{-\pi(2r+1)\over n} +O(n^{-2})  &\le &\sin\left({\pi (2r+1) x^*\over n}\right)
\le \sin \left({\pi \over 2r+1}\right) + O(n^{-1}).
\end{eqnarray*}
Thus,
$$
\lambda_{max}\ge \lambda_{x^*}\ge 1+(2r)^{-1} + O(n^{-1})
$$
and 
\be\lbl{ratio-ball}
{\lambda_{max}\over \lambda_{min}} \ge {3n^2 \over \pi^2 2r(r+1)} +O (n^{-1}).
\ee
\end{ex}

Estimate (\ref{ratio-ball}) shows that for any fixed $r\in\N$ the coupled system (\ref{coup})
may satisfy (\ref{ratio}) for small $n$ and fails to do so starting with some critical
network size $n^*(r)$ (see Fig.~\ref{f.Cay} \textbf{a}-\textbf{c}). For larger values of $r$, 
(\ref{coup}) satisfies (\ref{ratio}) for
larger $n$. In this respect, the present case is not different from the nearest-neighbor 
coupled networks, which we discussed in Example~\ref{ex.B1}. The situation changes
if we consider nonlocal coupling, i.e., $r=\lfloor \rho n \rfloor$ with the range of coupling
$\rho\in (0, 0.5)$. Then clearly for not too small $\rho$, one can have (\ref{ratio}) even
for $n\gg 1$.

\subsection{The complete, random, and quasirandom graphs}\lbl{sec.rand}
The complete graph on $n$ nodes $K_n=(V,E)$ is defined by 
$V=[n]$ and  $E=\{ ij: 1\le i<j\le n\}$.
\begin{lem}\lbl{lem.Kn}
The normalized graph Laplacian of $K_n$ has a simple zero EV
and the EV $1+(n-1)^{-1}$ of multiplicity $n-1$.
\end{lem}
\pf Since $K_n$ is a connected $(n-1)$-regular graph, the zero EV of
$L(K_n)$ is simple, and $\1_n$ is the corresponding eigenvector (cf.~(\ref{diagonal})).
Next, note that any nonzero vector $v=(v_1,v_2,\dots,v_n)$ orthogonal to $\1_n$ is
an eigenvector of $L(K_n)$. Indeed, suppose  $v_i\neq 0$ for some $i\in [n]$.
Then 
$$
(L(K_n)v)_i=v_i-{1\over n-1} \sum_{j=2}^n v_j=\left(1+{1\over n-1}\right) v_i,
$$
where the orthogonality condition $\sum_{j=1}^nv_j=0$ was used.\\
$\qed$

Therefore, the synchronization condition (\ref{ratio}) always  holds
for $\Gamma=K_n,$ $n\ge 2$. Furthermore, for sufficiently large $n$, the synchronization interval 
(\ref{domain}) is nonempty, and is practically independent of $n$. The windows of synchronization
shown in Fig.~\ref{f.K} \textbf{a}-\textbf{c} do not change as the size of the graph increases
from $n=10$ in Fig.~\ref{f.K}\textbf{a} to $n=100$ in Fig.~\ref{f.K}\textbf{c}. Below we consider (\ref{coup}) on several
graphs that approximate $K_n$ and, like $K_n$, exhibit good synchronization properties. 
\begin{figure}
\begin{center}
{\bf a}\hspace{0.1 cm}\includegraphics[height=1.8in,width=2.0in]{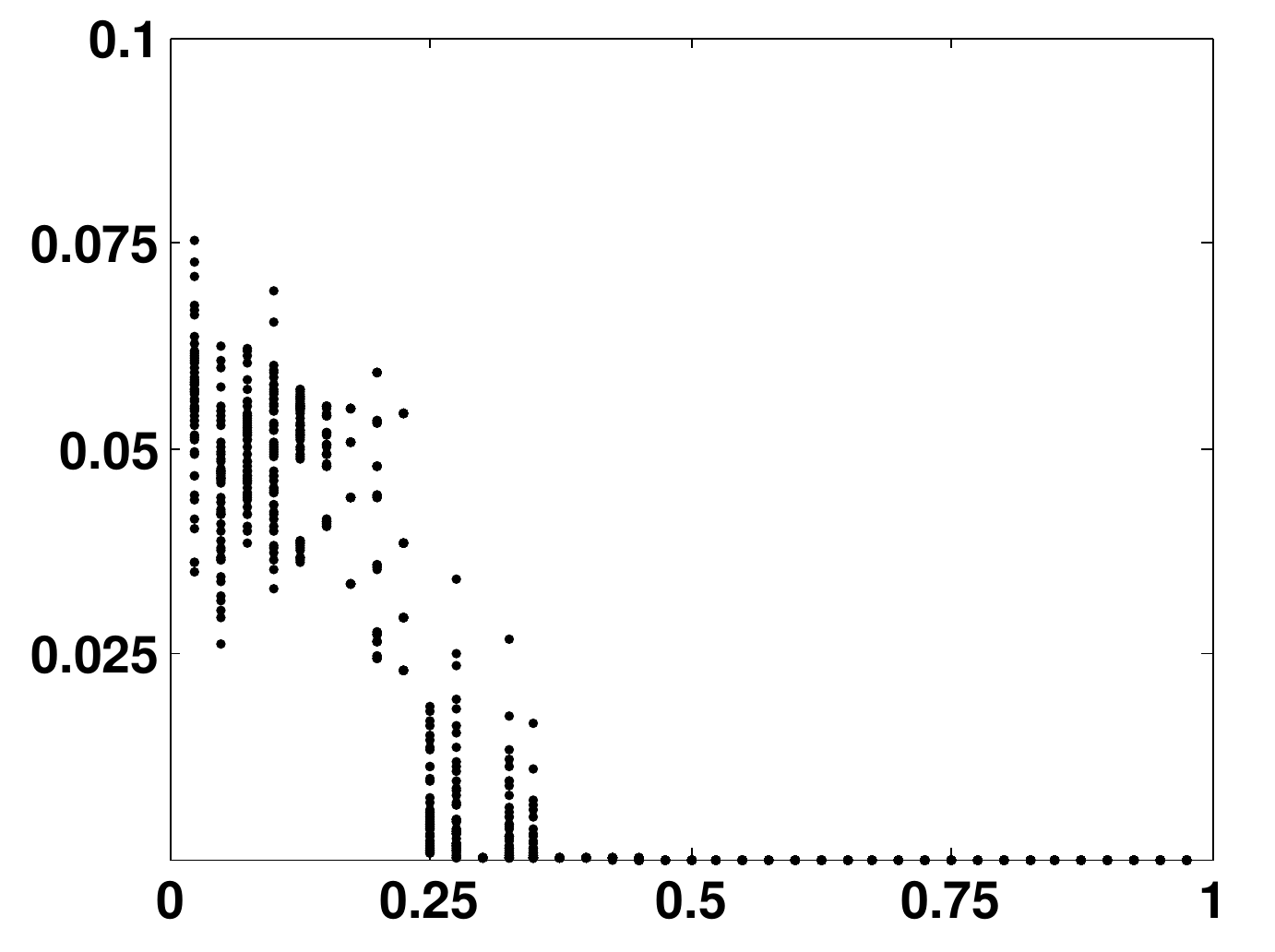}
{\bf b}\hspace{0.1 cm}\includegraphics[height=1.8in,width=2.0in]{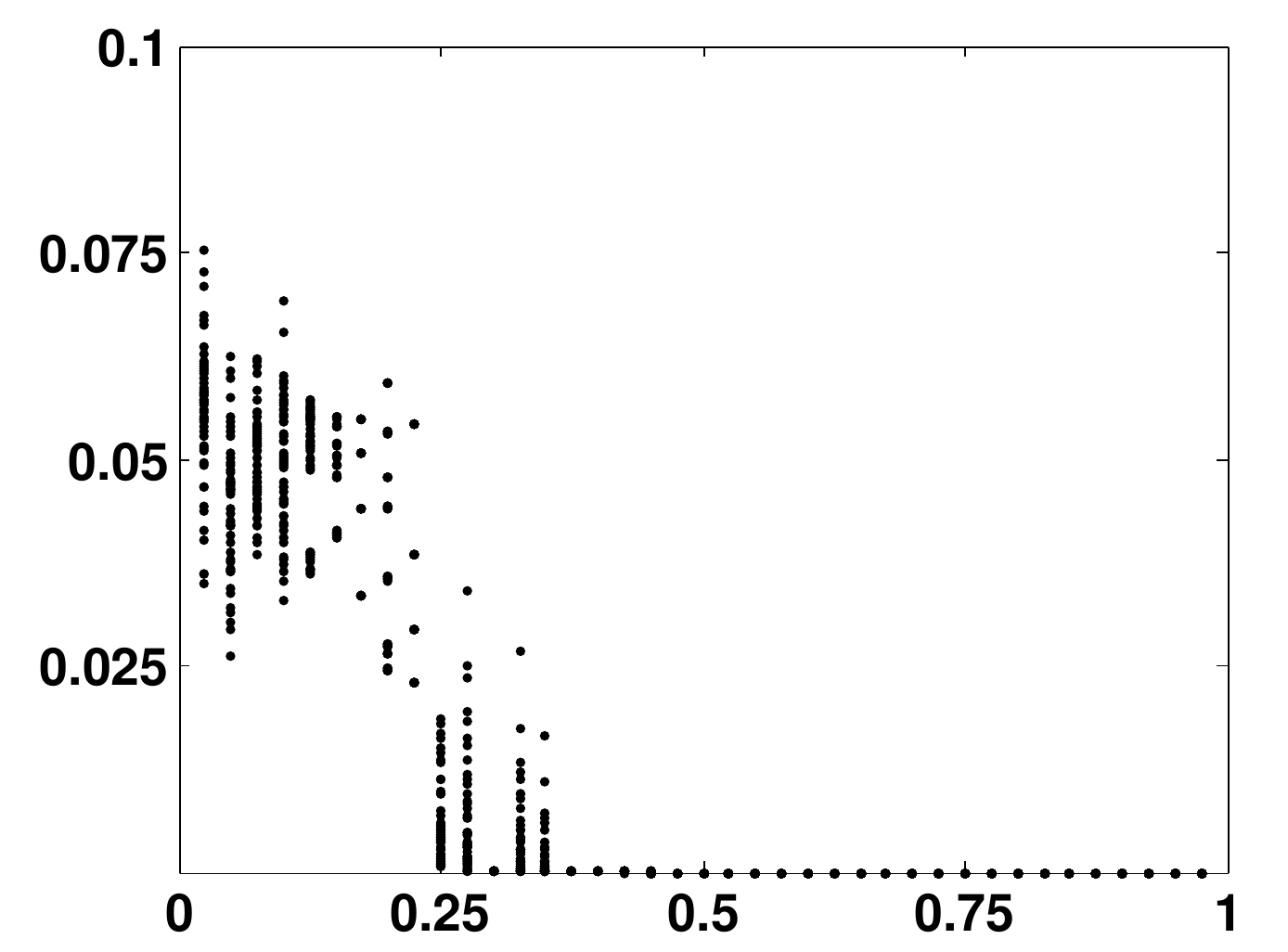}
{\bf c}\hspace{0.1 cm}\includegraphics[height=1.8in,width=2.0in]{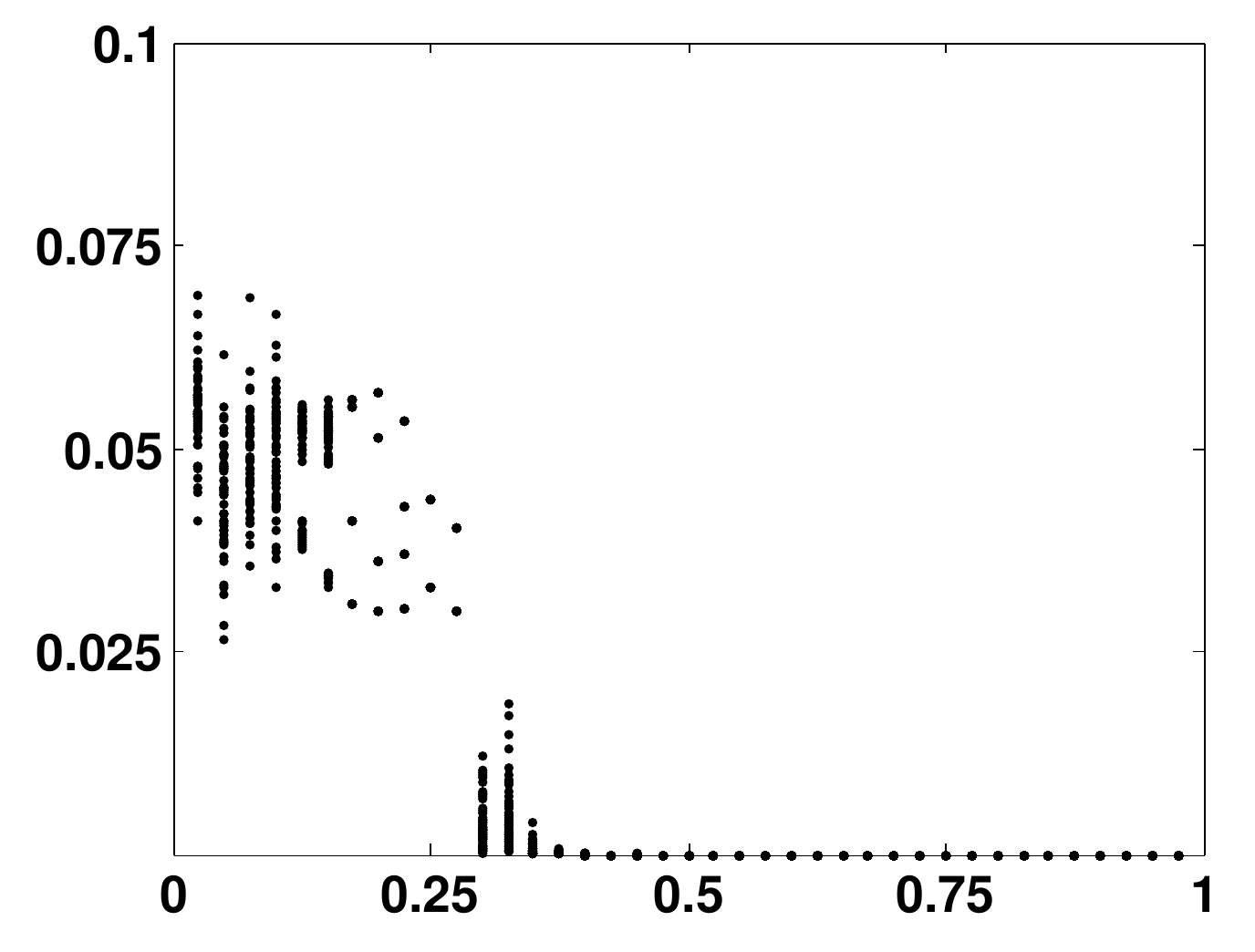}\\
{\bf d}\hspace{0.1 cm}\includegraphics[height=1.8in,width=2.0in]{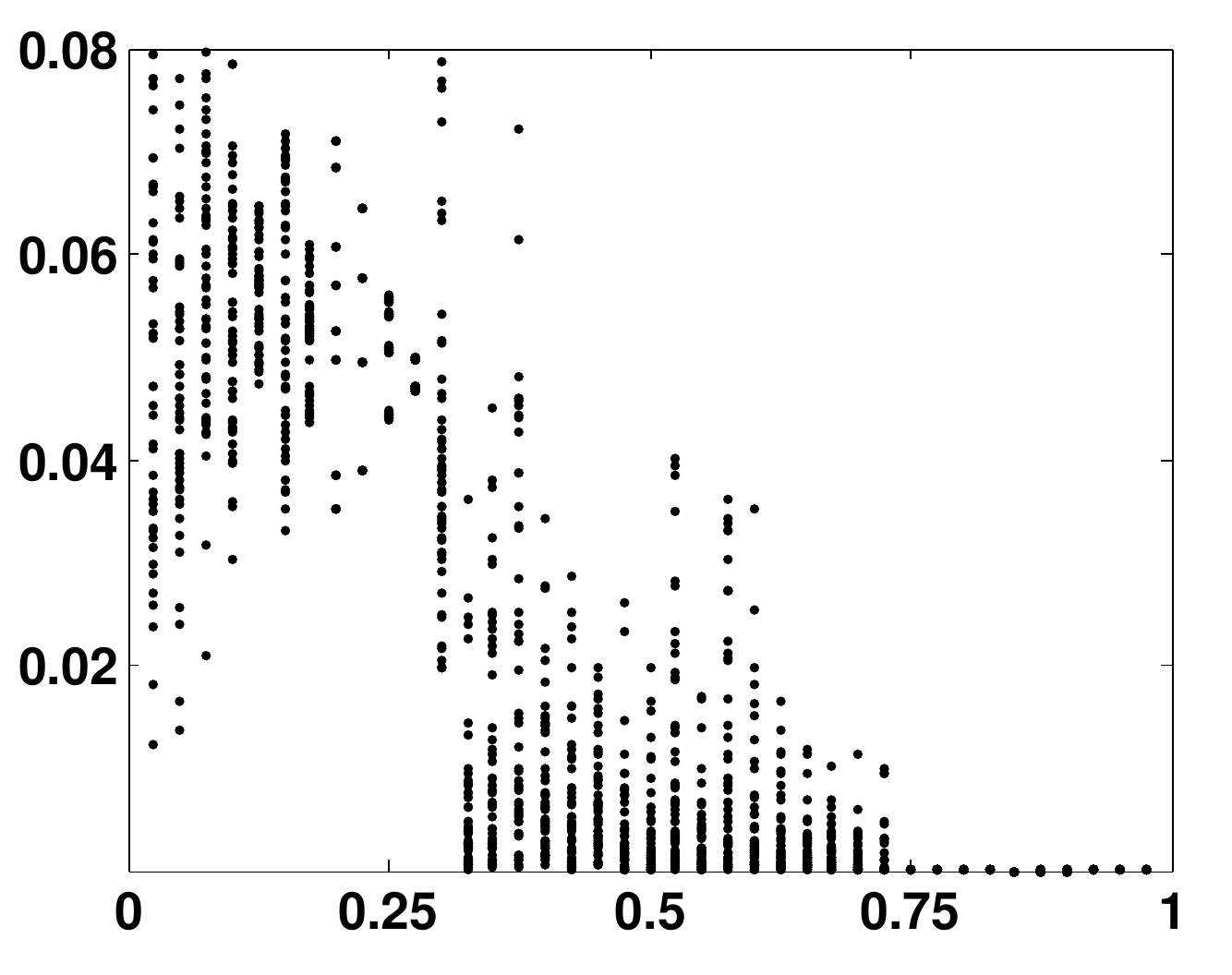}
{\bf e}\hspace{0.1 cm}\includegraphics[height=1.8in,width=2.0in]{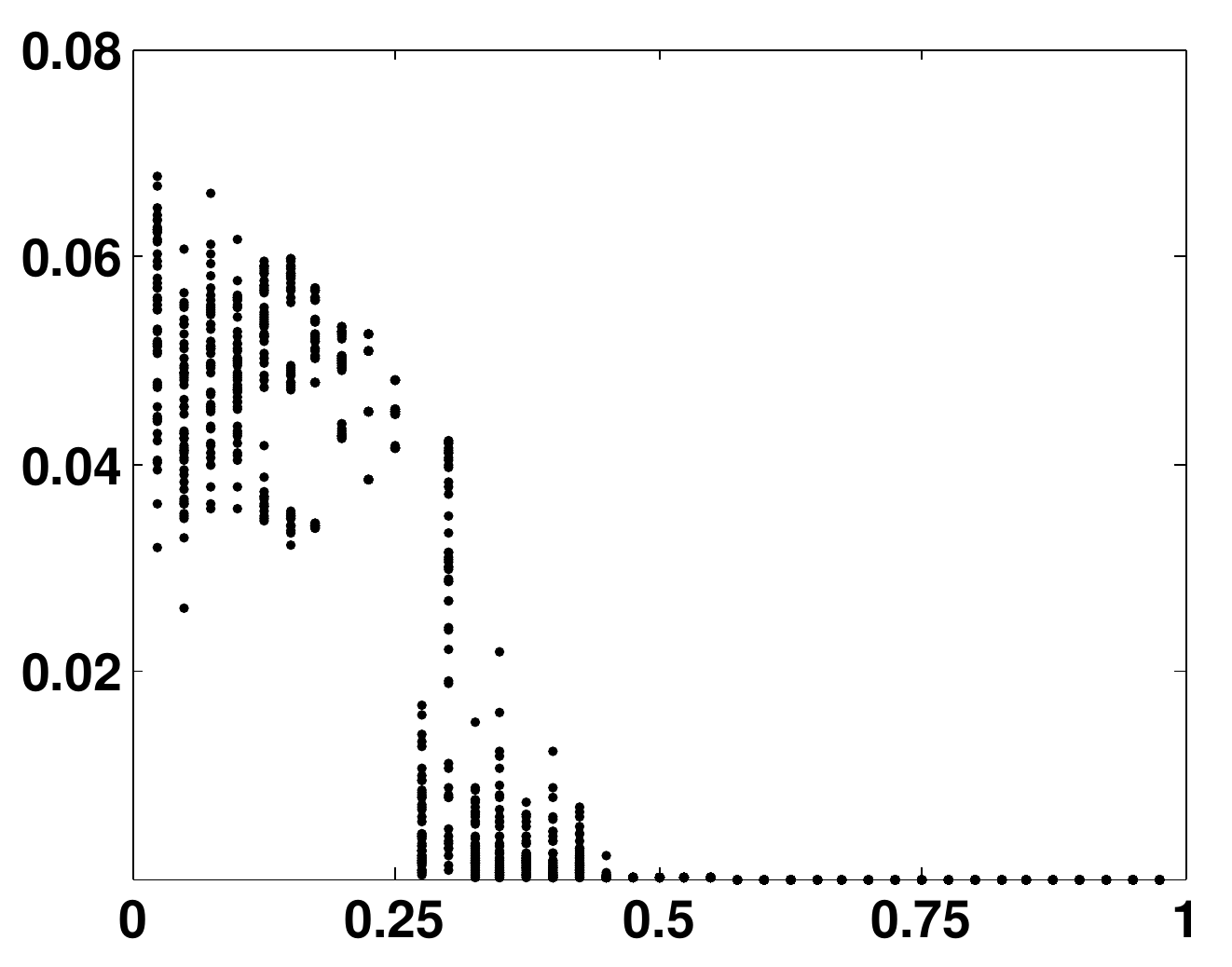}
{\bf f}\hspace{0.1 cm}\includegraphics[height=1.8in,width=2.0in]{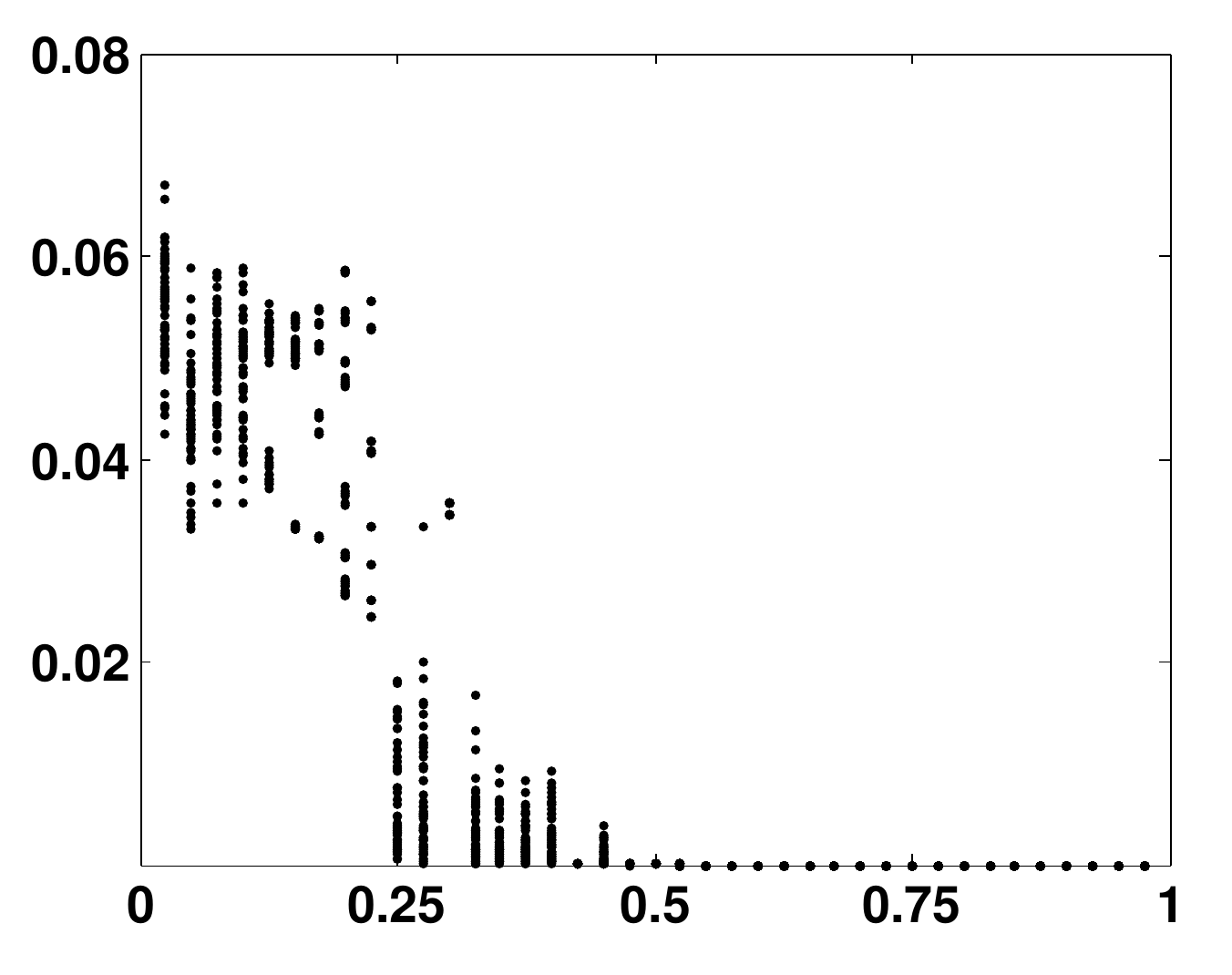}\\
{\bf g}\hspace{0.1 cm}\includegraphics[height=1.8in,width=2.0in]{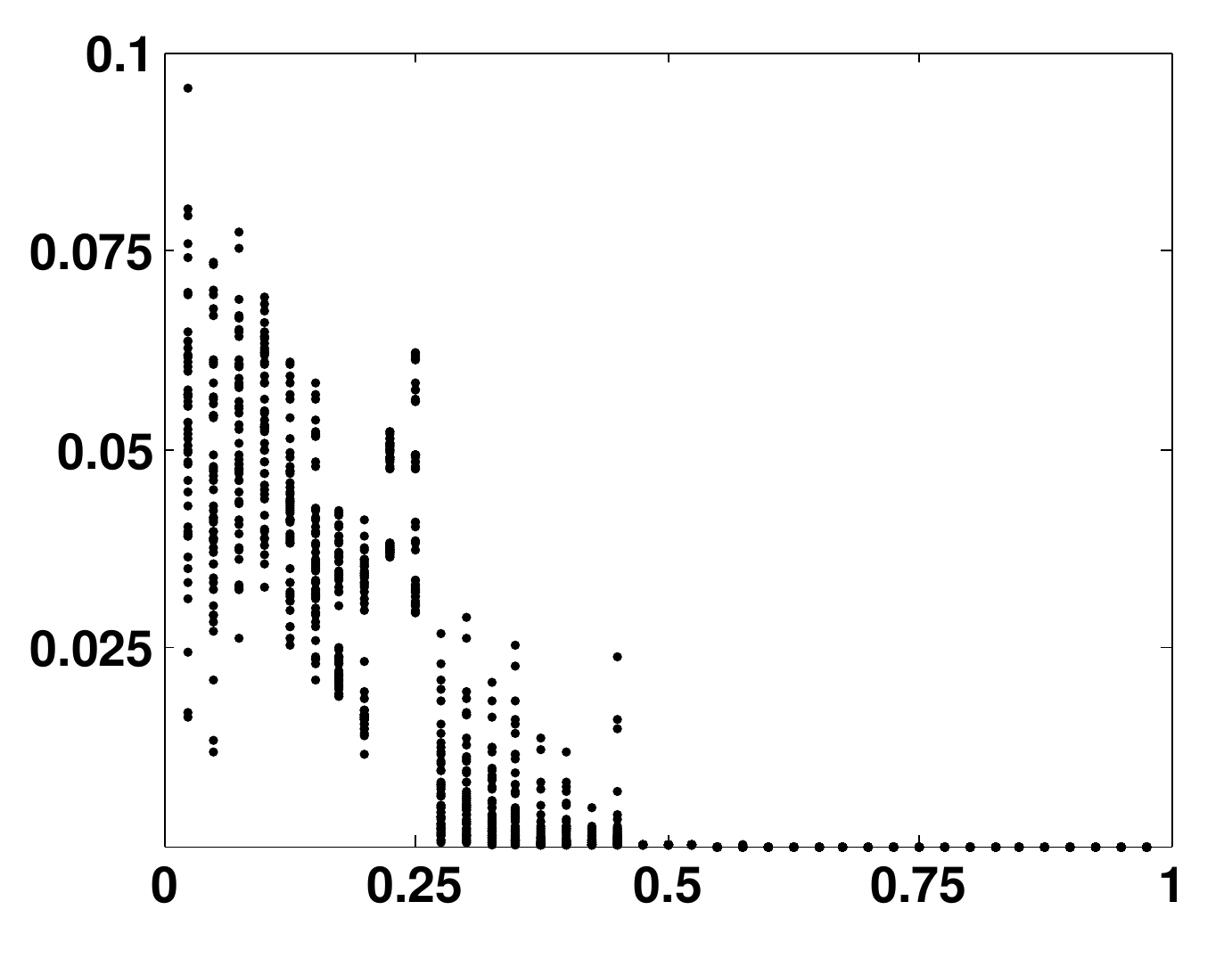}
{\bf h}\hspace{0.1 cm}\includegraphics[height=1.8in,width=2.0in]{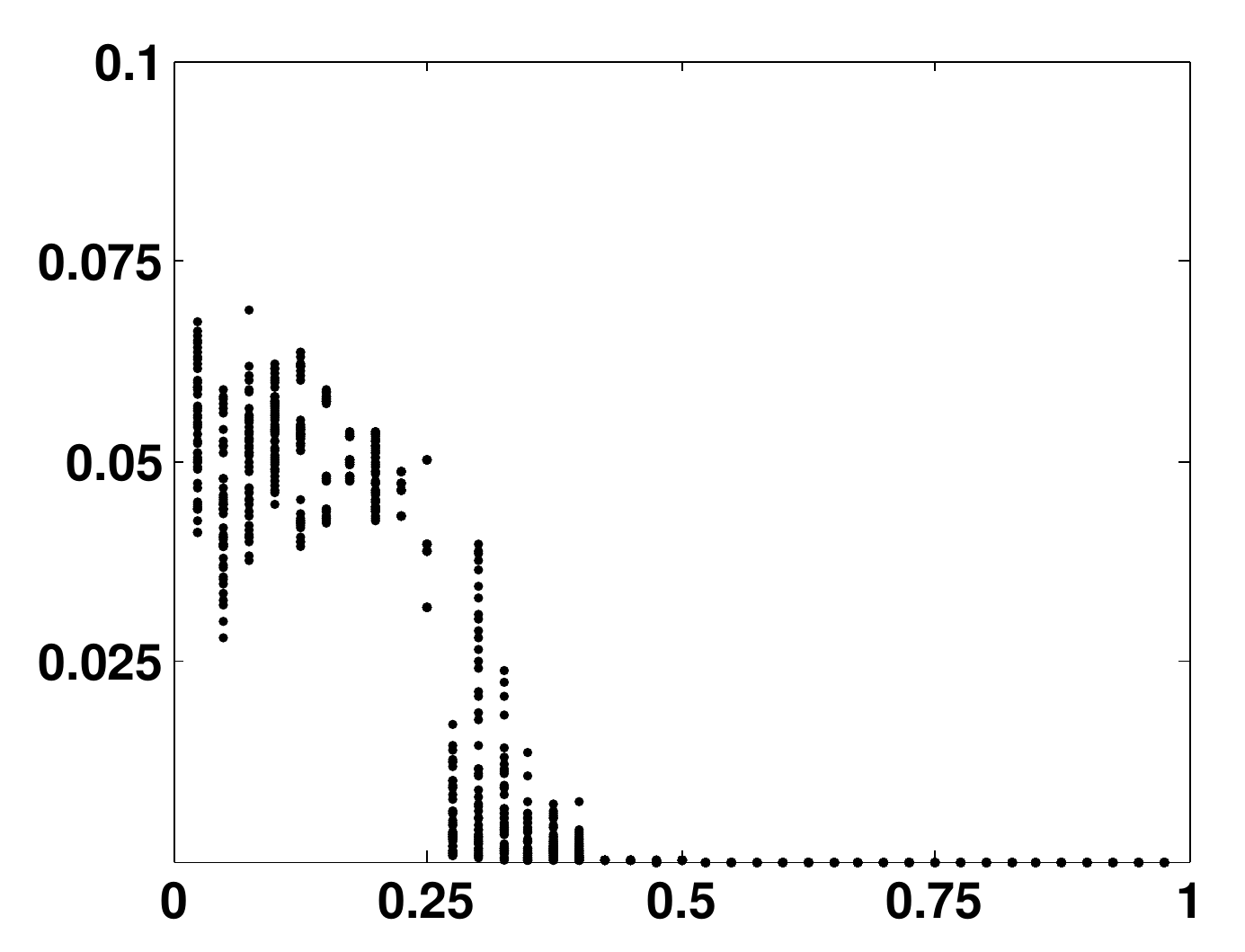}
{\bf i}\hspace{0.1 cm}\includegraphics[height=1.8in,width=2.0in]{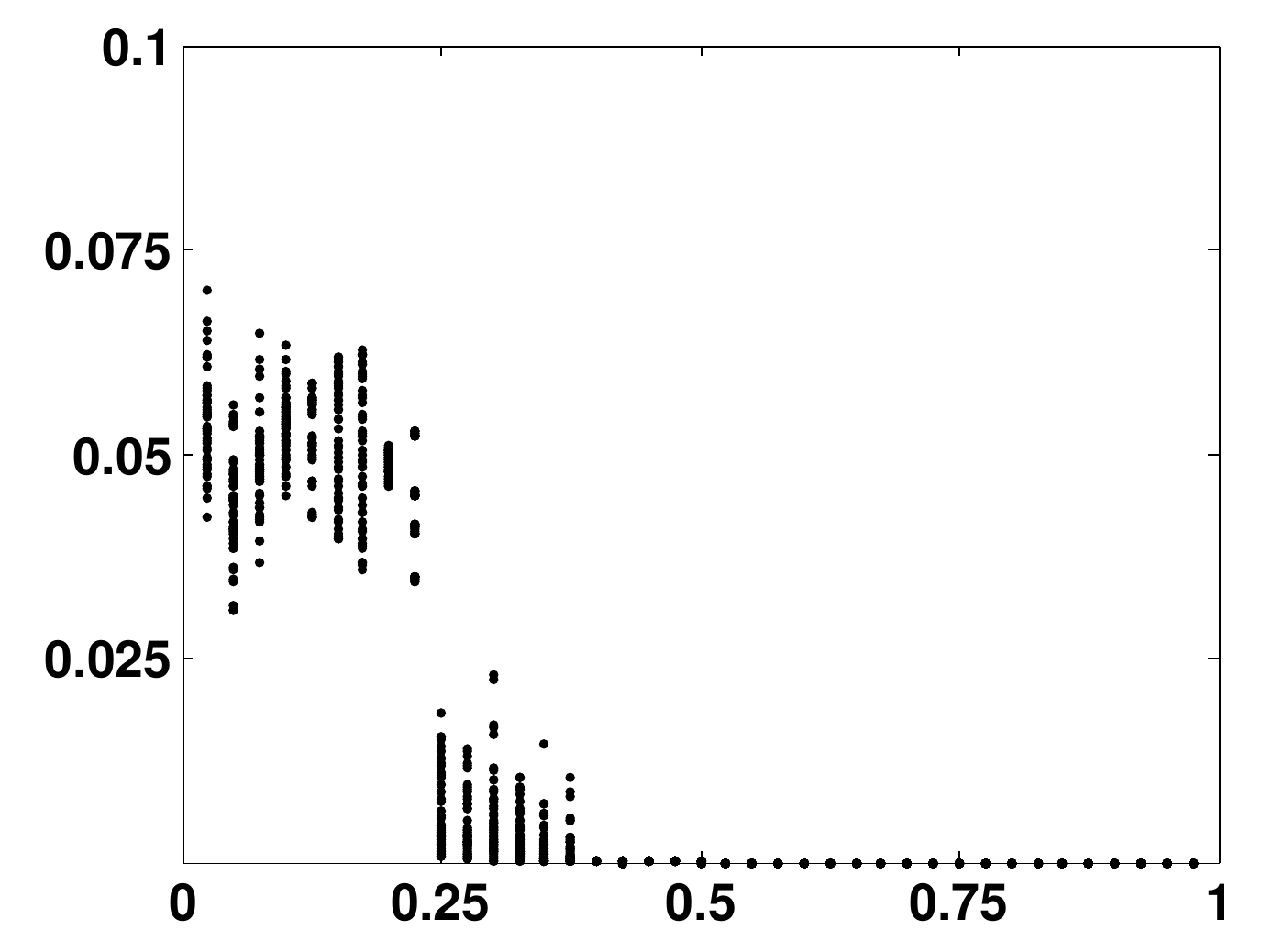}
\end{center}
\caption{ The range of the variance of  typical trajectories of (\ref{coup}) 
on $K_n$ (\textbf{a}-\textbf{c}), $G(n,1/2)$ (\textbf{d}-\textbf{f}), and
Paley (\textbf{g}-\textbf{i})  graphs plotted versus the coupling strength
parameter $\epsilon$. The graph sizes are
{\bf a})  $n=10,$ 
\textbf{b)} $n=50,$ \textbf{c)} $n=100,$ \textbf{d)} $n=10,$  
\textbf{e)} $n=50,$  \textbf{f)} $n=100$, \textbf{g)} $n=13,$  
\textbf{h)} $n=53,$ and  \textbf{i)} $n=101$.
}
\lbl{f.K}
\end{figure}

We begin with the classical Erd{\H{o}}s-R{\'e}nyi (ER) model of random graphs
\cite{Bollobas-MGT}. 
In this model, the edge between
every pair $ij, 1\le i<j\le n,$ is inserted with probability $p$. The decision for a given pair is 
made independently from the decisions on other pairs. The resultant
graph on $n$ nodes is denoted by $G(n,p)$ (see Fig.~\ref{f.1}\textbf{c}). The  nonzero EVs of 
$G(n,p)$ satisfy
\be\lbl{localize-rnd}
\max_{i\ge 2} |1-\lambda_i | \le O(n^{-1/2})
\ee
almost surely (cf.~\cite{ChuLu03, Oli10}). Therefore, from the synchronization viewpoint ER
 random graphs are as 
good as complete graphs. They also have considerably fewer edges than $K_n$, because
the expected degree of edges in $G(n,p)$ is $pn(n-1)/2$.

The EV concentration results in \cite{ChuLu03} apply to some other
random graph models including
power-law graphs (see also \cite{ChuRad11}). Specifically, let
$G(w)$ be a random graph  on $n$ nodes with a specified expected
degree sequence $w=(w_1,w_2,\dots, w_n)$ \cite{ChuLu02}. Chung-Lu random graphs on a vertex set
$V=[n]$ can be constructed by inserting an edge between $i$ and $j$ from $V$ with probability
$p_{ij}=w_iw_j \left(\sum_{k=1}^n w_k\right)^{-1}$. Here, we assume $\max_{i\in V} w_i^2\le \sum_{k=1}^n w_k$
and allow for loops. Further, for a given power law exponent $\beta$, maximum degree $m$,
and average degree $d$, one can generate a  power-law graph by setting (cf. \cite{ChuRad11})
$$
w_i=c (i+i_0)^{-1/\beta-1}, \; 1\le i\le n,
$$
where 
$$
c={\beta-2\over \beta -1} d n ^{1/\beta -1} \quad \mbox{and}\quad  i_0=\left( {d(\beta -2)\over m (\beta -1)}\right)^{\beta-1}.
$$
For Chung-Lu power-law graphs, Theorem~9 in \cite{ChuRad11} then guarantees that
all positive EVs are localized
$$
\left| \lambda_k(L)-1\right|=o(1), \; k\ge 2,
$$
with probability $1-o(1)$ provided $\min_{i\in V} w_i \gg \ln n$.
In Fig.~\ref{f.PL} we present numerical results for the coupled system 
(\ref{coup}) on Chung-Lu power-law graphs for two values of the expected degree, $d$.
For both values of $d$, Fig.~\ref{f.PL} shows large windows of synchronization.

\begin{figure}
\begin{center}
{\bf a}\hspace{0.1 cm}\includegraphics[height=1.8in,width=2.0in]{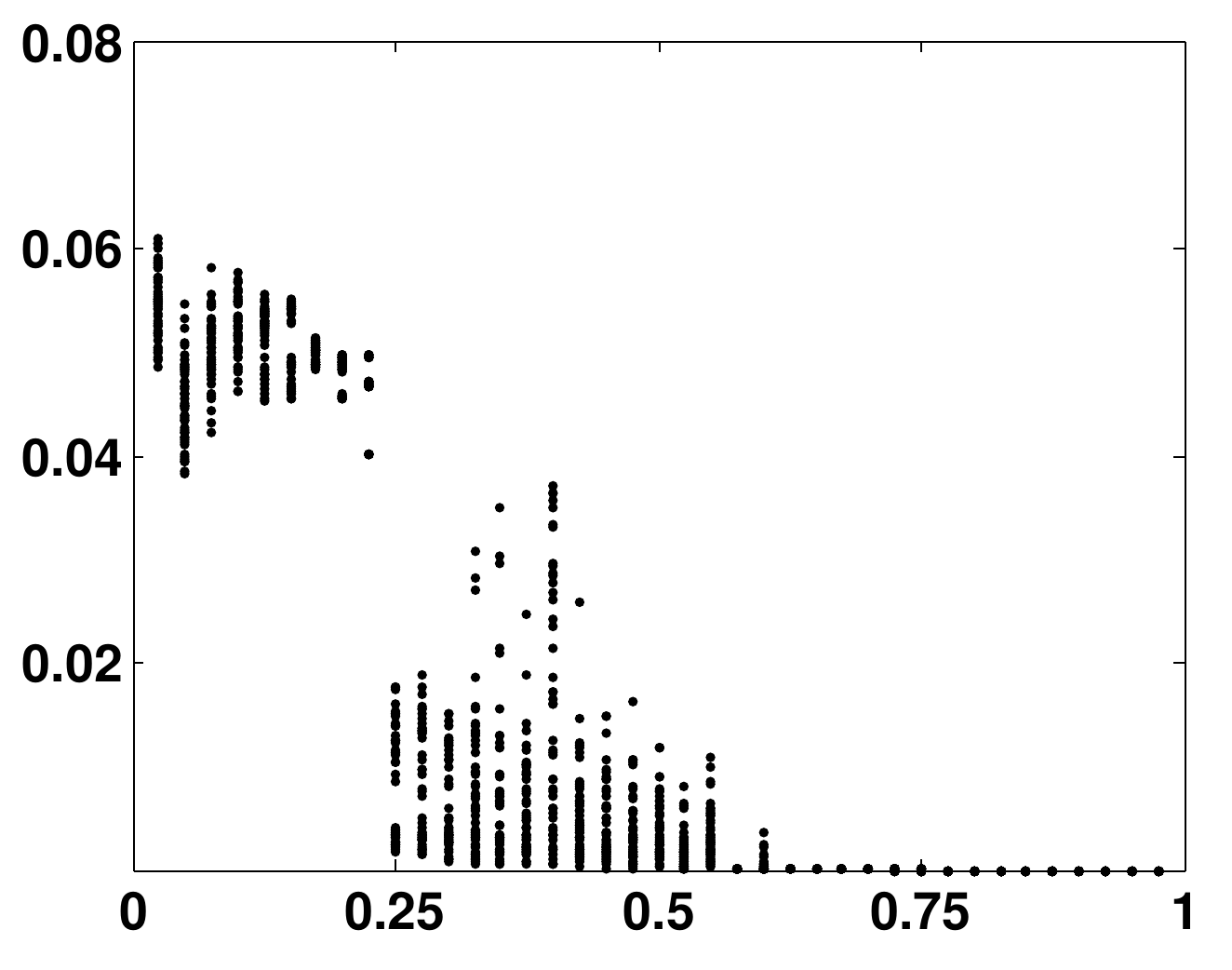}
{\bf b}\hspace{0.1 cm}\includegraphics[height=1.8in,width=2.0in]{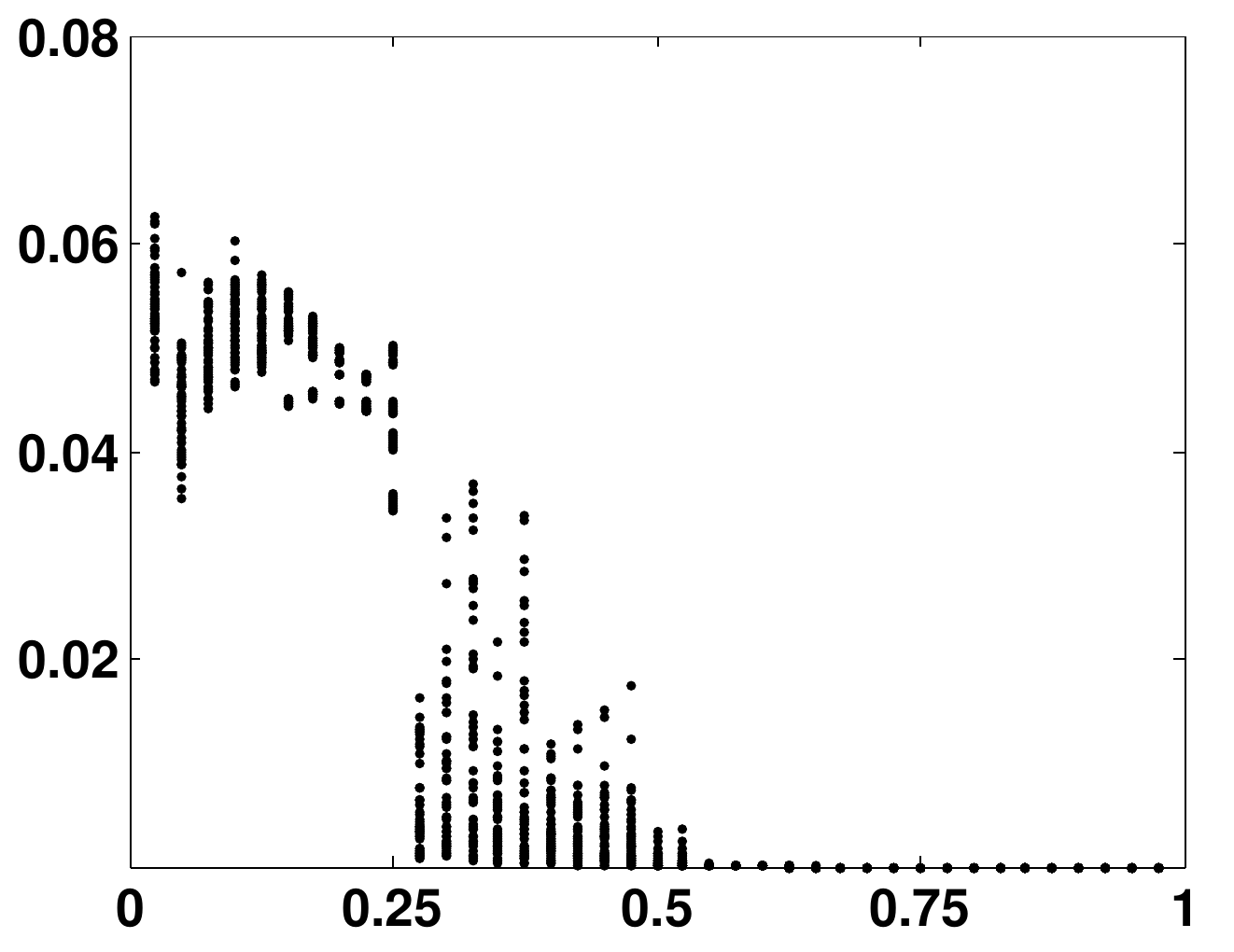}
\end{center}
\caption{The range of the variance of  typical trajectories of (\ref{coup}) 
on the power-law graphs on $200$ nodes.  
The degree distribution of the graphs used in all three plots is $\sim k^{-3}$. 
The average degree is \textbf{a}) $d=20$ and 
\textbf{b}) $d=30$.
}
\lbl{f.PL}
\end{figure}

Next, we turn to quasirandom graphs, which share many combinatorial properties  with  the  
ER graphs $G(n,p)$ (cf.~\cite{ChuGra88,KriSud06}). Below, we restrict to the case $p=1/2$ and consider
only $d$-regular graphs. There are several equivalent properties that define a quasirandom graph 
$\Gamma=(E,V)$ \cite{ChuGra88}. For instance, for any $S\subset V,\; (|V|=n)$,
the number of edges connecting the nodes in $S$ is 
\be\lbl{edge-asymp}
{1\over 4} |S|^2+O(n^2),
\ee
exactly as for the ER graph $G(n,1/2)$. If $\Gamma$ is a $d$-regular graph then (\ref{edge-asymp})
is equivalent to the following localization property of the nonzero EVs of the normailized graph
Laplacian (cf.~Theorem~9.3.1, \cite{AloSpe-PMethod})
\be\lbl{localize-QR}
\left|\lambda_i-1\right|=o(1), \; i=2,3,\dots,n.
\ee
Hence, for coupled systems on quasirandom graphs synchronization condition (\ref{ratio}) holds 
for sufficiently large $n$ and, as in the case of the complete graph, the interval of synchronization 
(\ref{domain}) to leading order is independent of $n$ for $n\gg 1$ 
(see Fig.~\ref{f.K} \textbf{d}-\textbf{f}). To illustrate synchronization properties of (\ref{coup})
on quasirandom graphs, we use Paley graphs, which are defined as follows.
 
\begin{ex}\lbl{ex.Paley}
Let  $n= 1\pmod 4$ be a prime and denote 
$$
\Z^\times_n=\Z_n/\{0\} \quad\mbox{and}\quad Q_n=\{ x^2 \pmod{n}:\; x\in \Z_n^\times\}.
$$  
$Q_n$ is viewed as a set (not multiset, i.e., each element has multiplicity $1$). Then $Q_n$ is 
a symmetric subset of $Z_n^\times$ and $|Q_n|=2^{-1}(n-1)$ (cf. \cite[Lemma~7.22]{KreSha11}).
$P_n=\Cay(Z_n, Q_n)$ is called a Paley graph \cite{KreSha11}.

The nonzero EVs of $L(P_n)$ can be computed explicitly from (\ref{EV-C}), using properties of 
the Gauss sum (cf.~\cite{KreSha11}):
 \be\lbl{EV-Pn}
\lambda_x(P_n)=
\left\{\begin{array}{ll}
1 + {1-\sqrt{n}\over n-1}, & x \mbox{ is a QR}~\pmod{n}, \\
1 + {1+\sqrt{n}\over n-1}, & x \mbox{ is not a QR}~\pmod{n},
\end{array}
 \right.
\ee
where $x\in\Z_n^\times$.
\end{ex}

The numerical results for the coupled system (\ref{coup}) on Paley graphs show good
agreement with the analytical estimates. The synchronization windows computed numerically
for (\ref{coup}) on Paley graphs are as large as those for the complete graphs of the same
size (see Fig.~\ref{f.K} \textbf{g}-\textbf{i}). Overall, the variance plots for (\ref{coup}) on 
the complete, ER, and Paley graphs in Fig.~\ref{f.K} have a great degree  of similarity.

\subsection{Expanders}
The complete, random, and quasirandom graphs endow the coupled system (\ref{coup})
with good synchronization properties but at a price of having $O(n^2)$ edges in a graph
of size $n$. In this subsection, 
we will review certain families of graphs with much fewer edges, which nonetheless
approximate complete graphs and promote synchronization in coupled systems like
(\ref{coup}).

Let $\Gamma_n=(V(\Gamma_n),E(\Gamma_n)), n\in\N,$ be a family of $d$-regular connected graphs.
Suppose $V(\Gamma_n)=[n]$ and denote the EVs of 
$L(\Gamma_n)$ by
$$
0=\lambda_1<\lambda_2\le \dots\le\lambda_n.
$$
By Gershgorin's Theorem \cite{HornJohn-Matrix}, $\lambda_n\le 2d$. Therefore,
the most likely way for the synchronization condition (\ref{ratio})  to break down 
for large $n$ is  through the second EV $\lambda_2(\Gamma_n)$ approaching $0$.
This leads us to consider families of graphs, for which the second EV
remains bounded away from zero
\be\lbl{gap}
\lambda_2(\Gamma_n)\ge c_1>0.
\ee
 Such families of graphs are called expanders \cite{Sar04}.

\begin{df}\lbl{df.expander}\cite{ReiVad02, KreSha11}
A family of $d$-regular graphs $\Gamma_n=(V(\Gamma_n), E(\Gamma_n)$
($|V(\Gamma_n)|\to \infty$ as $n\to \infty$) is called an expander family if the 
Cheeger constant 
\be\lbl{Cheeger}
h(\Gamma_n)=\min_F {|\partial F|\over \min\{|F|, |V(\Gamma_n)/F|\}}
\ge c_2>0 \;
\forall n,
\ee
 where the minimum is taken over all proper subsets of $V(\Gamma_n)$, $F\neq \emptyset$,
and $\partial F$ stands for the set of edges connecting the nodes in $F$ with the nodes
in the complement of $F$, $V(\Gamma_n)/F$.
\end{df}
The Cheeger constant $h(\Gamma)$ quantifies the connectivity of $\Gamma$. The uniform 
bound on $h(\Gamma_n)$ in (\ref{Cheeger}) guarantees that $\Gamma_n$ remain well-connected
even as  $|V(\Gamma_n)|$ grows without a bound. The combinatorial condition
(\ref{Cheeger}) is equivalent to the spectral one (\ref{gap}) (cf.~\cite{AloMil85, Alo86}),
which we arrived at in the context of synchronization. Therefore, synchronizability in large
networks is directly related to connectivity: the better connected large networks are, the  more likely
they
to satisfy the synchronization condition (\ref{ratio}). The connectivity is better in graphs 
with larger second EV $\lambda_2$ \cite{Fied73}. There is an upper bound on $\lambda_2(\Gamma_n)$
(cf.~\cite{HooLin06}):
\be\lbl{AlBop}
\limsup_{n\to\infty} \lambda_2(\Gamma_n) \le d-2\sqrt{d-1}.
\ee
A family of graphs $\{\Gamma_n\}$ is called Ramanujan if 
$$
\lambda_2(\Gamma_n) \ge d-2\sqrt{d-1}.
$$
Ramanujan graphs exhibit the optimal asymptotics of $\lambda_2(\Gamma_n)$ and,
in this sense, possess the best possible connectivity.
By inspecting several common families of $d$-regular graphs, such as the nearest-neighbor
graphs, one can see that typically the second EV tends to $0$ rapidly as $n\to\infty$ 
(see, e.g., (\ref{lmin-Br})).
In fact, any family of Cayley graphs of bounded degree on an abelian group
can not be an expander family \cite[\S 4.3]{KreSha11}.  
Nontheless, a typical family of random
$d$-regular graphs ($d\ge 3$) is very close to be Ramanujan with high probability
 as shown in \cite{Fri08}.
The deterministic constructions of expander families 
are also available,  albeit they 
rely on sophisticated algebraic techniques (cf.,~\cite{Mar88, LubPhi88, ReiVad02}).
\begin{figure}
\begin{center}
{\bf a}\hspace{0.1 cm}\includegraphics[height=1.8in,width=2.0in]{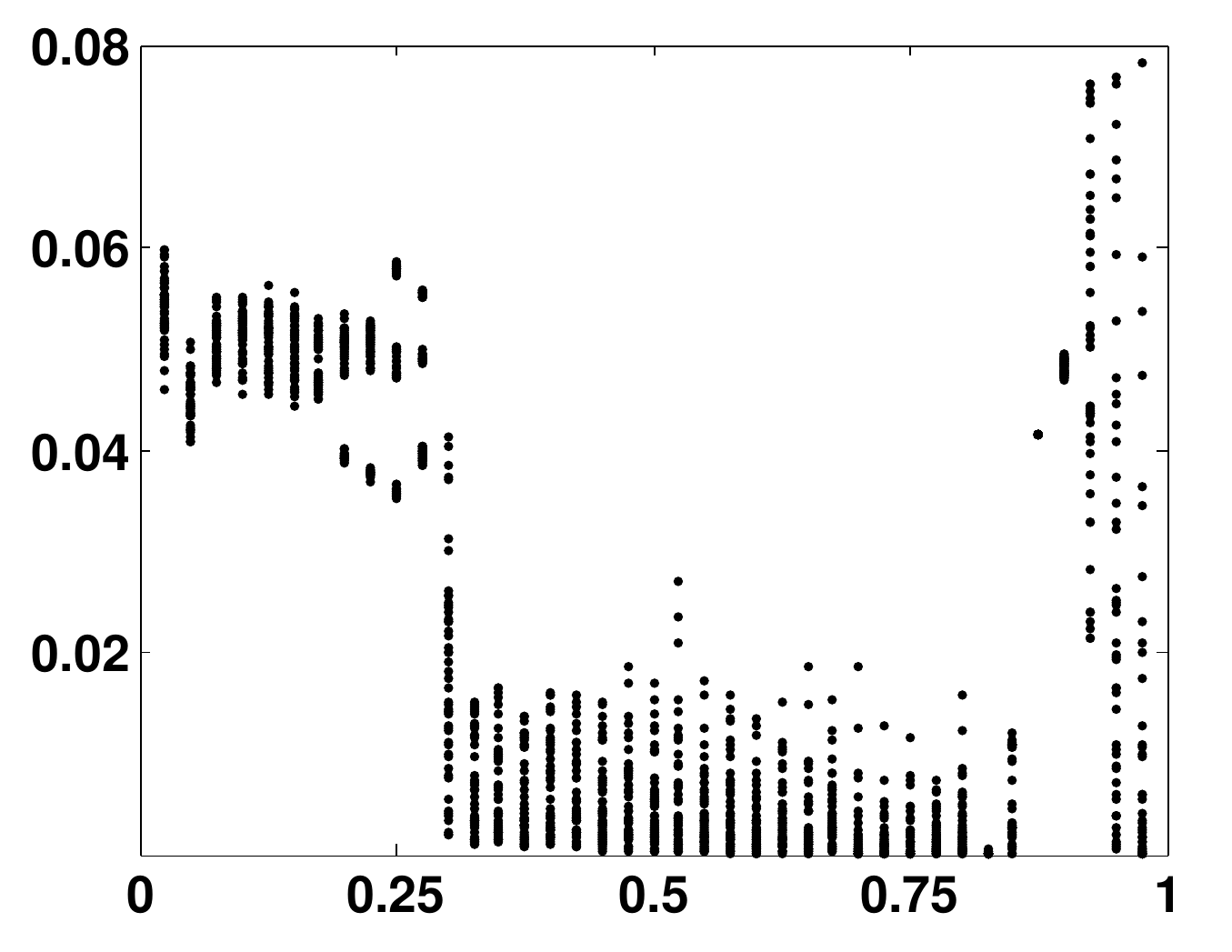}
{\bf b}\hspace{0.1 cm}\includegraphics[height=1.8in,width=2.0in]{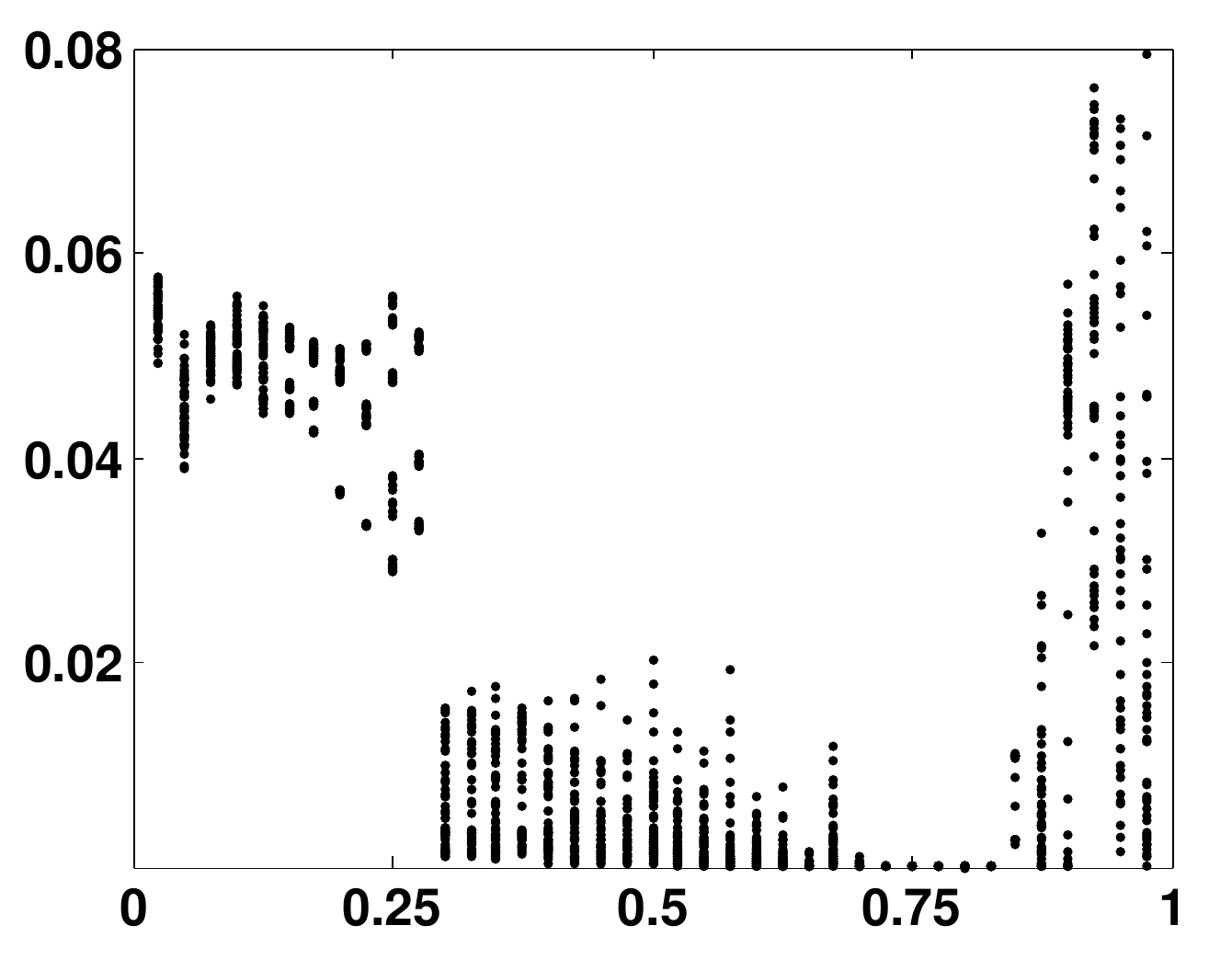}
{\bf c}\hspace{0.1 cm}\includegraphics[height=1.8in,width=2.0in]{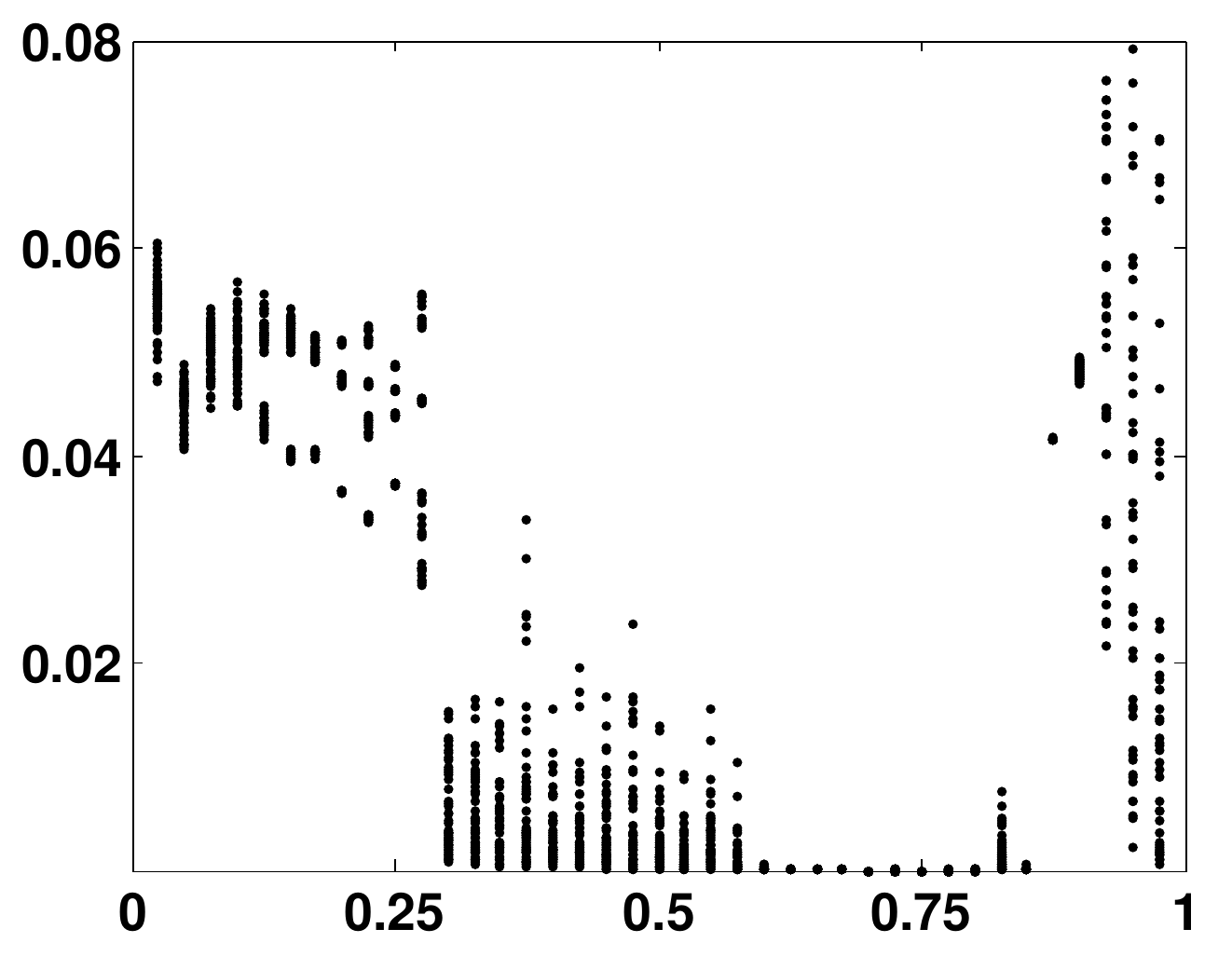}\\
{\bf d}\hspace{0.1 cm}\includegraphics[height=1.8in,width=2.0in]{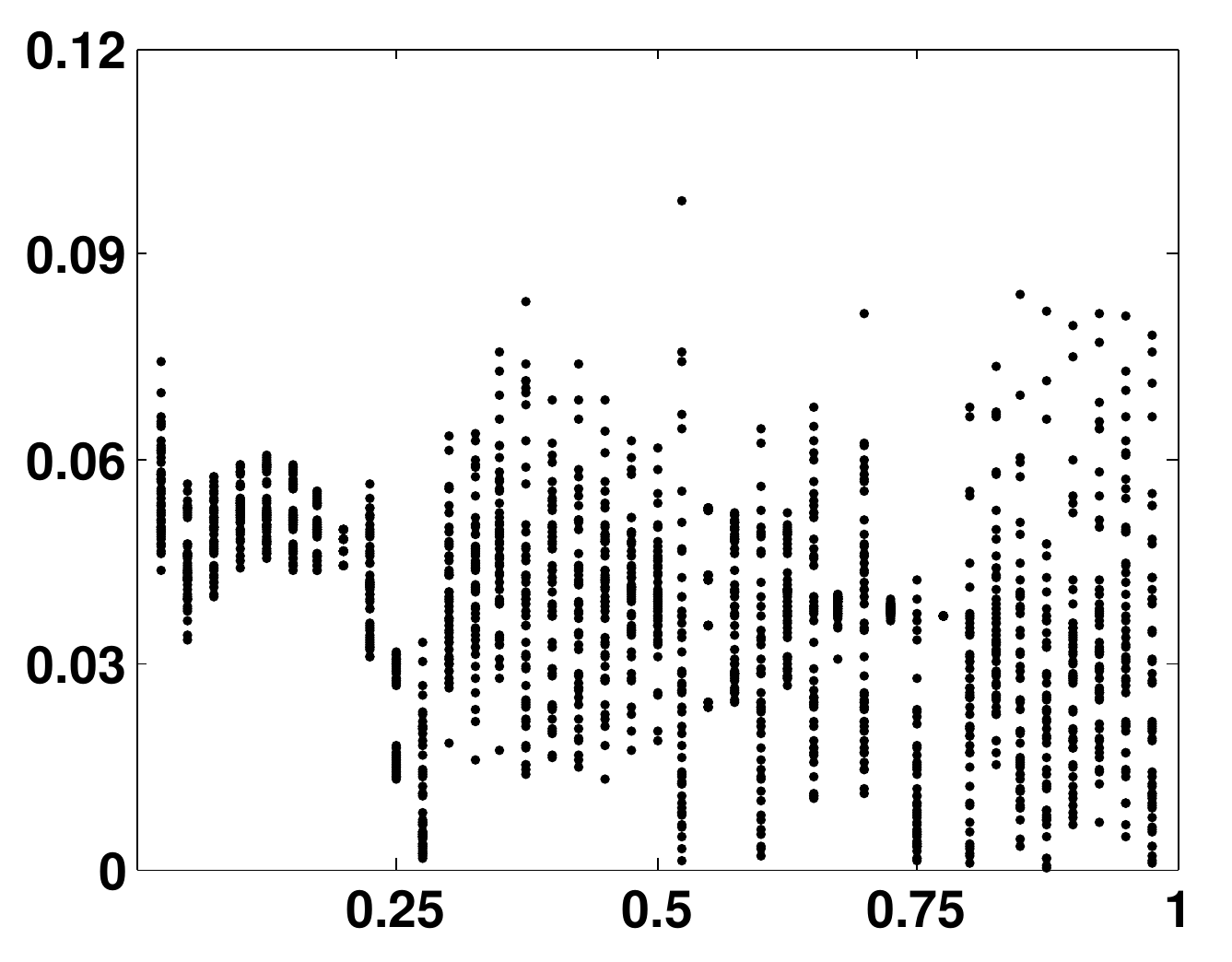}
{\bf e}\hspace{0.1 cm}\includegraphics[height=1.8in,width=2.0in]{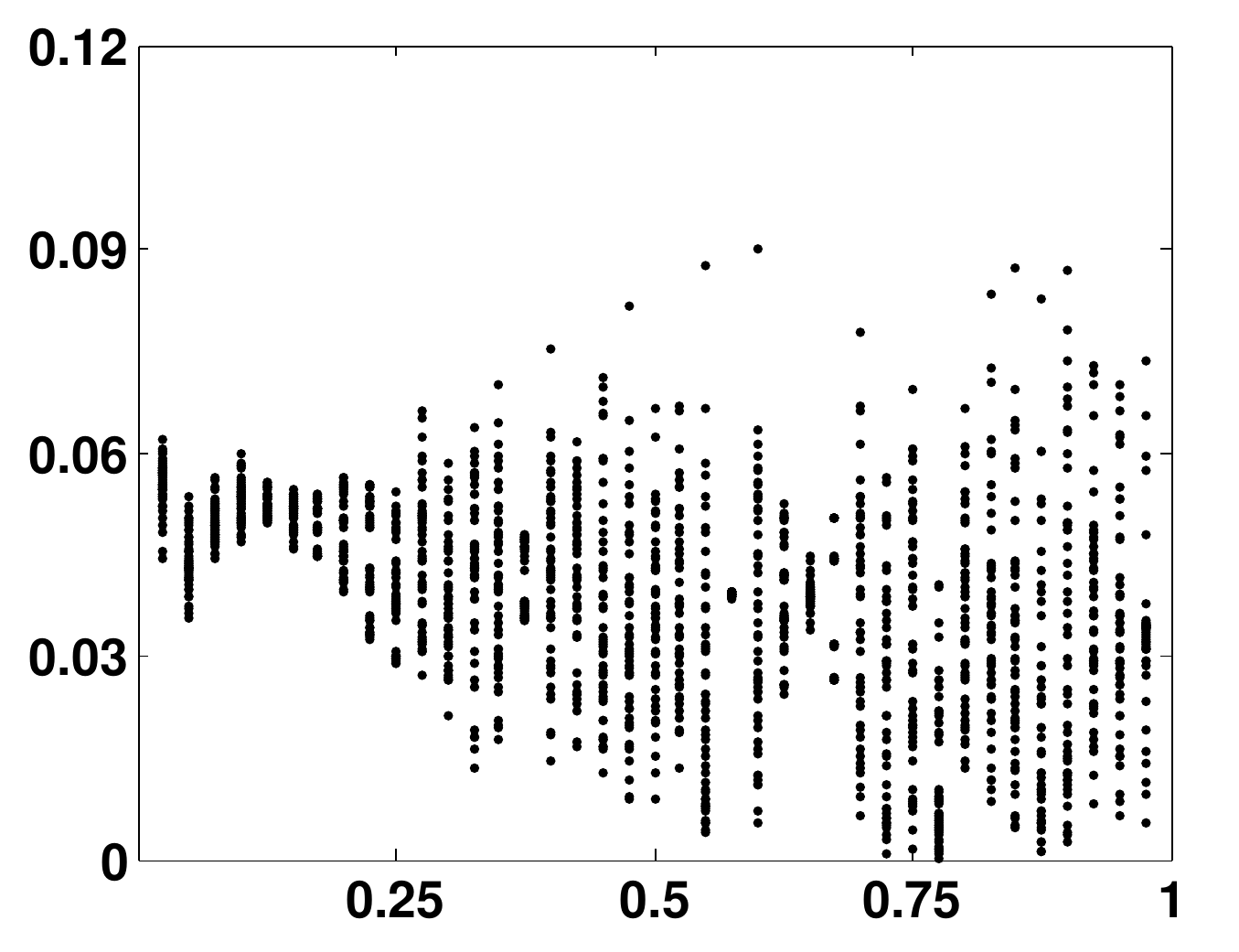}
{\bf f}\hspace{0.1 cm}\includegraphics[height=1.8in,width=2.0in]{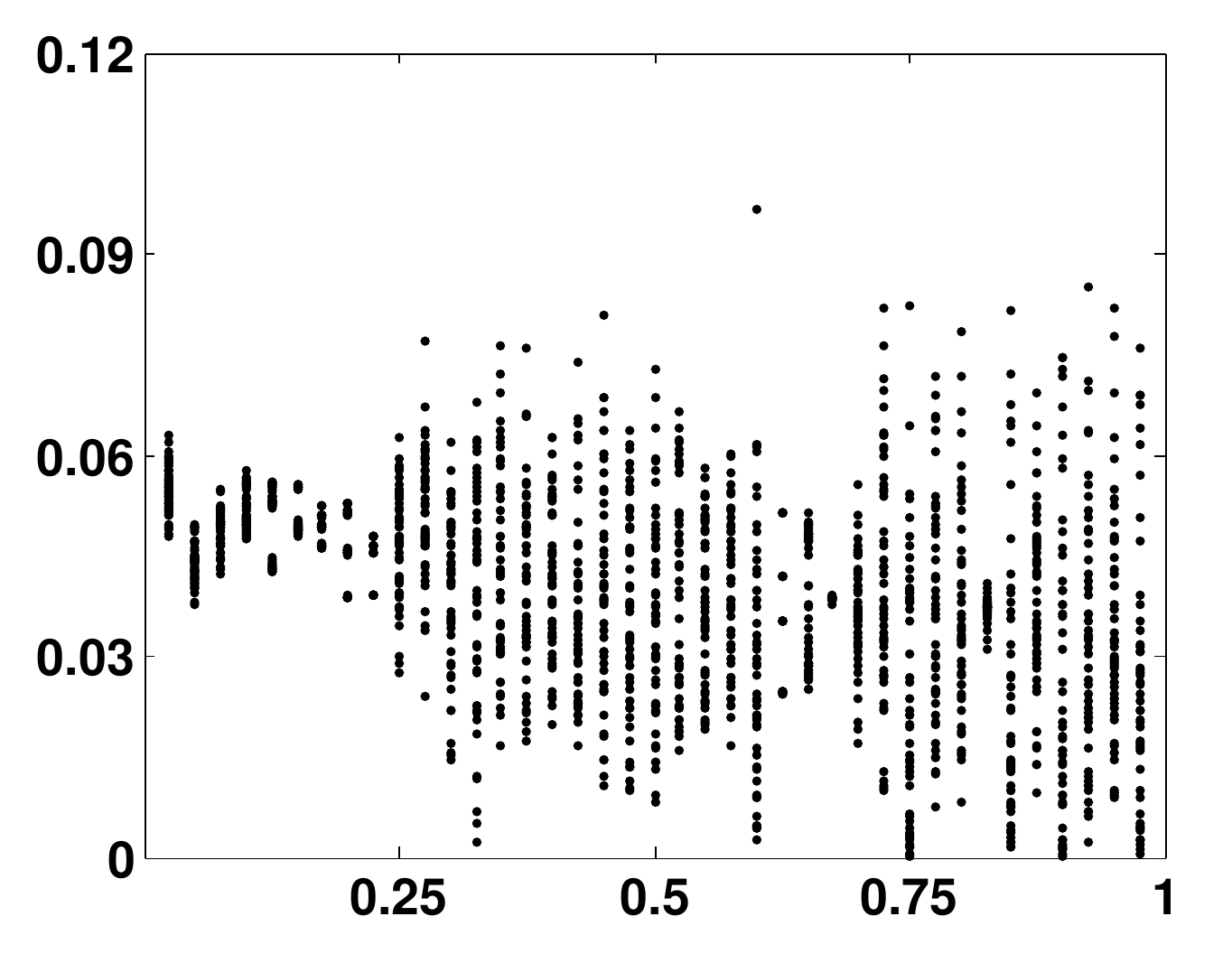}\\
\end{center}
\caption{The range of the variance of  typical trajectories of (\ref{coup}) 
on the random bipartite graphs (cf.~Example~\ref{ex.d-rnd}). The graphs used in these
simulations have $400$ nodes. The degrees are \textbf{a}) $10,$ \textbf{b}) $15,$
and \textbf{c}) $20.$ For comparison, in plots \textbf{d}-\textbf{f}, we included
numerical results for the coupled system on Cayley graphs with regular connections
$\Cay(Z_n, B(r))$ for \textbf{d}) $n=100, r=10$, \textbf{e}) $n=200, r=20$, and 
\textbf{f}) $n=400, r=40$.
}
\lbl{f.BP}
\end{figure}

To illustrate synchronization in coupled systems on expanders, we will use the following
family of random bipartite $d$-regular graphs.
\begin{ex}\lbl{ex.d-rnd}
Let $B_{2m,d}, m\ge 2$ be a bipartite graph on $2m$ vertices.
The edges are generated using the following algorithm:
\begin{enumerate}
\item
Let $p:[m]\to[m]$ be a random permutation.
In our numerical experiments, we used MATLAB function $\mathsf{randperm}$
to generate random permutations.
For $i\in [m]$, add edge $(i, m+p(i))$.
\item
Repeat step 1. $d-1$ times.
\end{enumerate}
\end{ex}
\begin{figure}
\begin{center}
{\bf a}\hspace{0.1 cm}\includegraphics[height=1.8in,width=2.0in]{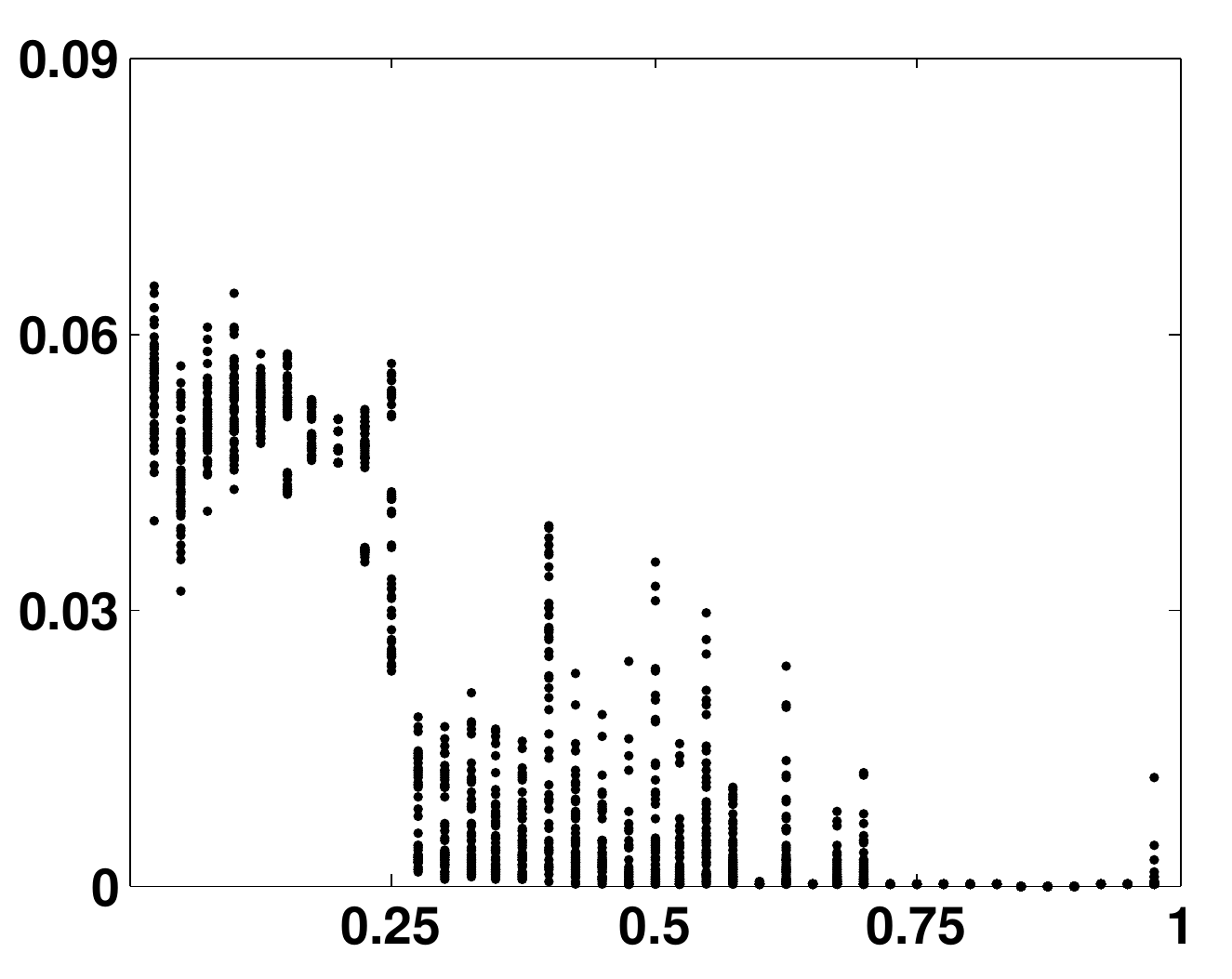}
{\bf b}\hspace{0.1 cm}\includegraphics[height=1.8in,width=2.0in]{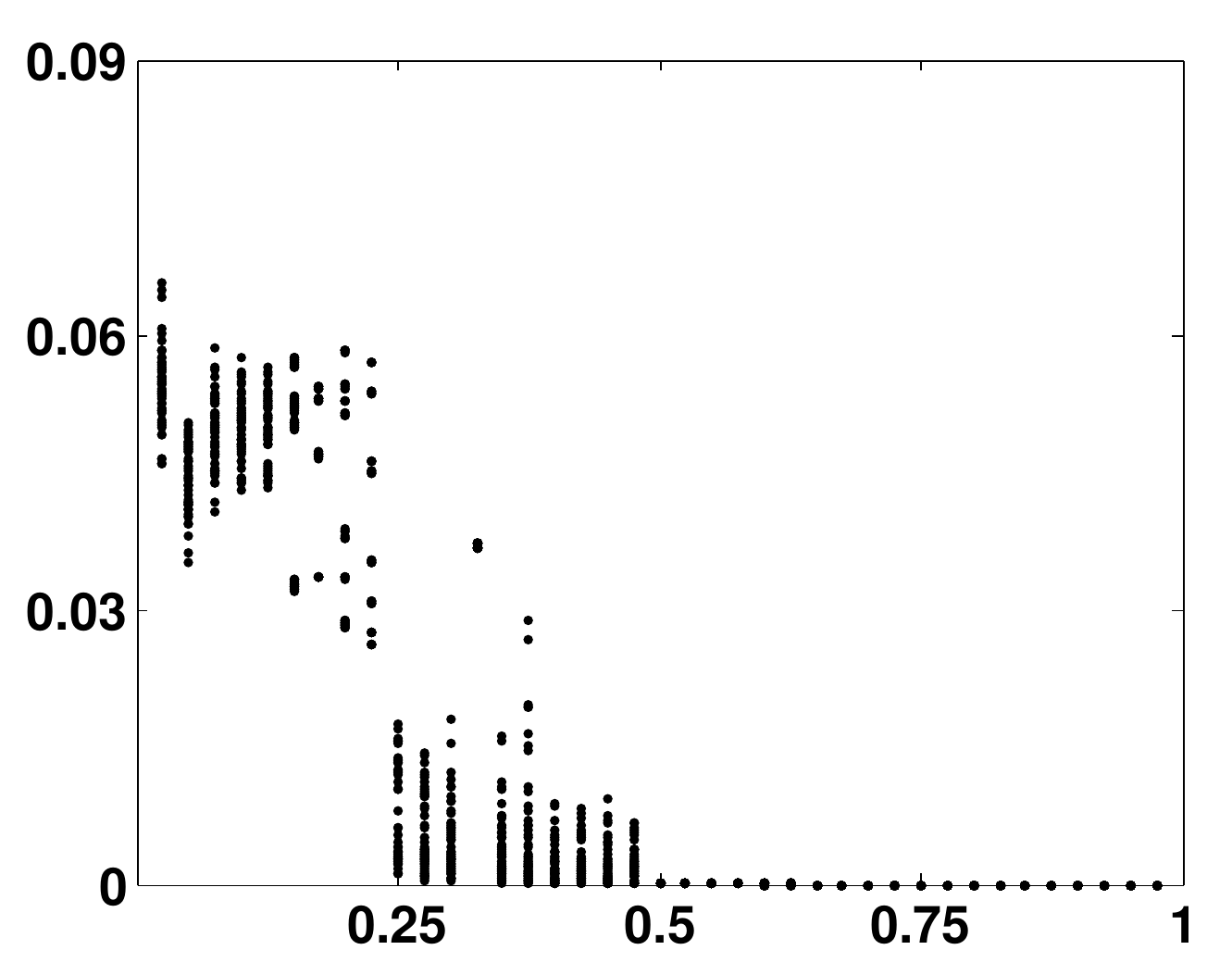}
{\bf c}\hspace{0.1 cm}\includegraphics[height=1.8in,width=2.0in]{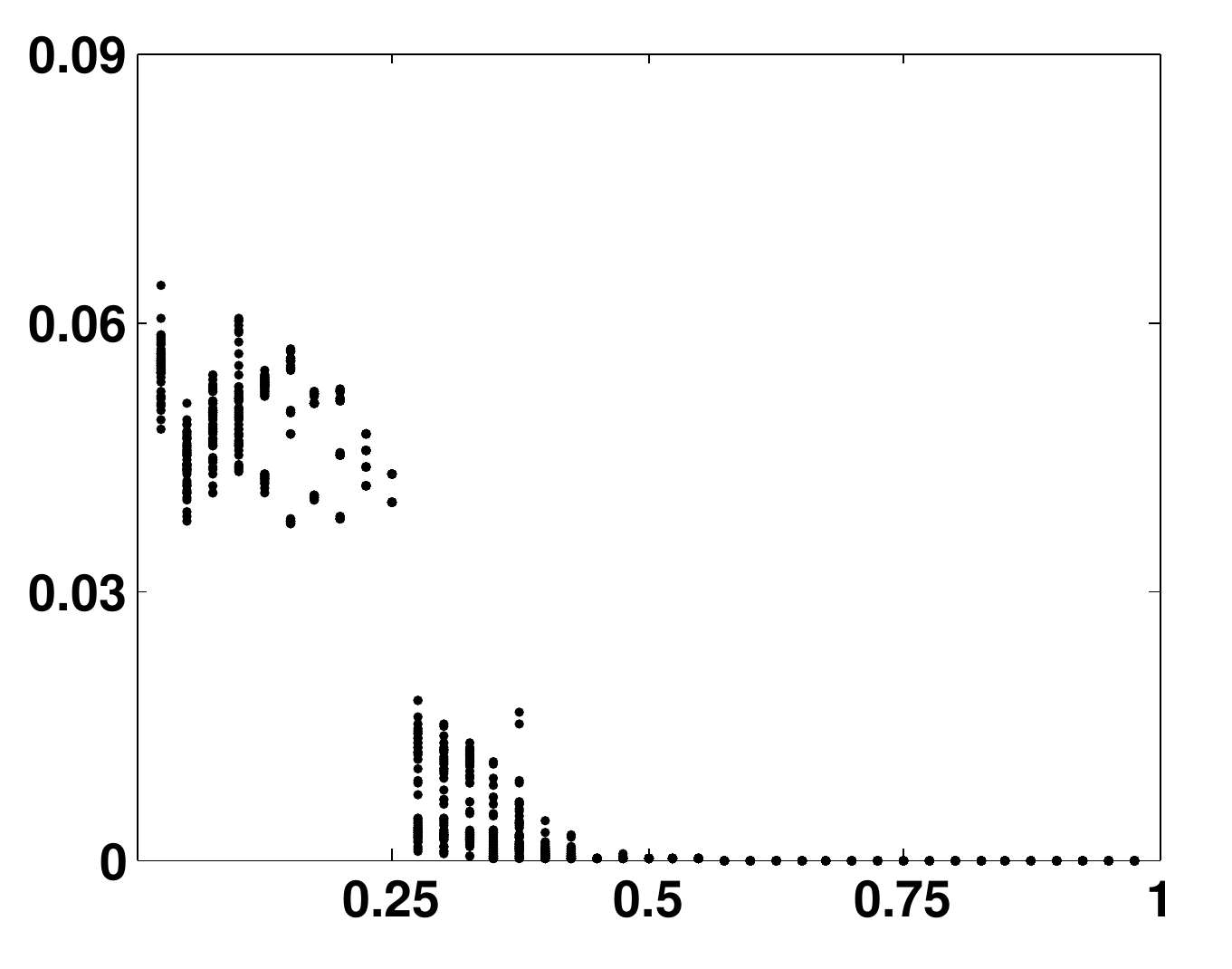}
\end{center}
\caption{The range of the variance of  typical trajectories of (\ref{coup}) 
on the random Cayley graph $\Cay(\Z_n, \pm R_k)$: \textbf{a}) $n=100$, $k=10$, 
\textbf{b}) $n=200$, $k=20$,  and  \textbf{c}) $n=400$, $k=40$.
}
\lbl{f.RCay}
\end{figure}

The numerical results in Fig.~\ref{f.BP} show that coupled system on $B_{400,15}$ already has a
notable window of synchronization (see Fig.~\ref{f.BP}\textbf{b}), which becomes
even bigger for $B_{400,20}$ (see Fig.~\ref{f.BP}\textbf{c}). 
Note that in the present case we do not have localization results for the EVs of $L$, 
as in the analysis of the
coupled system (\ref{coup}) on random and quasirandom graphs (cf.~(\ref{localize-rnd}), 
(\ref{localize-QR})), and we can only rely on the lower bound for $\lambda_2$ (\ref{gap}).
Therefore, for  chaotic dynamical systems on expanders, in general, we can not guarantee
that synchronization condition (\ref{ratio}) holds for all $n\gg 1$. Nonetheless, as can be
seen from the numerical results in Fig.~\ref{f.BP} \textbf{a}-\textbf{c}, one can achieve
synchronization on large expanders of relatively small degree. The numerical results
for the bipartite random graphs may be compared 
to the results for the Cayley graphs on $B(r)$ in Fig.~\ref{f.BP} \textbf{d}-\textbf{f}, which show that the 
coupled system (\ref{coup}) on smaller  graphs of the same degree but with regular connections 
remains rather far from synchrony. This is another illustration of the advantages of random connectivity
for synchronization.

In conclusion, we consider synchronization of the coupled system (\ref{coup}) on random Cayley graphs 
\newline
$\Cay(\Z_n, \pm R_k)$, where $R_k$ is a set of $k$ elements chosen from the uniform distribution
on $\Z_{\lfloor n/2\rfloor}$. Random Cayley graphs are expanders with high probability for $n\gg1 $ and 
$k=O(\log n)$ \cite{LanRus04}. The numerical results for the dynamical systems on random Cayley 
graphs in Fig.~\ref{f.RCay} show that these systems exhibit much better synchronization properties
than the systems on regular Cayley graphs of the same size  (see Fig.~\ref{f.BP} \textbf{d}-\textbf{f}).

\section{Discussion}
\setcounter{equation}{0}
Motivated by the linear stability analysis in \cite{JosJoy02}, in this
paper, we rigorously derived a sufficient condition for synchronization in systems of diffusively coupled
maps, whose dynamics can be chaotic. In contrast to the approach in \cite{JosJoy02}, we did not seek to
show asymptotic stability of spatially homogeneous solutions of (\ref{coup})
\be\lbl{homo}
\mathbf{x}_k= x_k\1_n, \; k=0,1,2,\dots,
\ee
where $\{x_k\}$ is a trajectory of the local system (\ref{loc}). Instead, we proved the asymptotic stability
of the invariant subspace of such solutions $\mathcal{D}$ (cf.~(\ref{diagonal})), in the sense that
every trajectory of (\ref{coup}) that started sufficiently close from $\mathcal{D}$ approaches $\mathcal{D}$
in forward time. Note that the asymptotic stability of $\mathcal{D}$ does not imply the asymptotic stability 
of individual trajectories (\ref{homo}). In particular, our approach circumvents the analysis of perturbations
in the tangential to $\mathcal{D}$ direction, which would be necessary for the justification of the linearization 
about (\ref{homo}). The latter problem is quite technical already in the case when local systems have 
stable dynamics \cite{Med11}. On the other hand, it is not clear that
for coupled chaotic systems one 
has sufficient control over perturbations
in the tangential direction, because it corresponds to a degenerate subspace of
the coupling matrix and the local dynamics are  sensitive to perturbations.

The sufficient condition for synchronization (\ref{ratio}) identifies the complete graph as an optimal
network organization for chaotic synchronization. Therefore, in the second part of this paper, we 
considered networks on graphs that approximate the complete graph. These include ER random graphs,
as well as quasirandom and Chung-Lu power-law graphs. We show that from synchronization point of view
these graphs are almost as good as the complete graph provided that the network is large enough.
Furthermore, expanders, which constitute a broader class of graphs, tend to have good synchronization
properties. As many of these graphs are constructed using random algorithms, our results highlight the 
advantages of random connectivity for synchronization.

\vskip 0.2cm
\noindent
{\bf Acknowledgements.}
This work was supported in part by the NSF grants  DMS 1109367 and DMS 1412066 (to GSM).

\providecommand{\bysame}{\leavevmode\hbox to3em{\hrulefill}\thinspace}
\providecommand{\MR}{\relax\ifhmode\unskip\space\fi MR }
\providecommand{\MRhref}[2]{%
  \href{http://www.ams.org/mathscinet-getitem?mr=#1}{#2}
}
\providecommand{\href}[2]{#2}

\end{document}